\documentclass[letterpaper,twocolumn,10pt]{article}
\usepackage{usenix2019_v3}

\usepackage{tikz}
\usepackage{amsmath}
\usepackage{amssymb}
\usetikzlibrary{positioning,decorations.pathreplacing,shapes}
\usepackage{hyperref}
\usepackage{cleveref}
\usepackage{todonotes}
\usepackage{enumitem}

\usepackage{minted}
\usepackage{caption}
\newcommand{\heading}[1]{\vspace{5pt}\noindent{\textbf{#1.}}}
\usepackage{xspace}
\usepackage{booktabs} 
\usepackage{graphicx}
\usepackage{multirow}

\definecolor{lavenderblush}{rgb}{1.0, 0.94, 0.96}
\definecolor{lightcyan}{rgb}{0.88, 1.0, 1.0}

\definecolor{varColor}{rgb}{0.72, 0.2, 0.64}
\newcommand{\variable}[1]{\color{varColor} \mathsf{#1}}
\newcommand{\concat}{\,\|\,}

\newcommand{\secref}[1]{\S \ref{#1}}

\newcommand\replicate[2]{%
  \ifnum#1>0%
    #2%
    \expandafter\replicate\expandafter{\number\numexpr#1-1}{#2}%
  \fi%
}%

\begin{document}

\date{}

\title{\Large \bf Security and Privacy Analysis of Tile's Location Tracking Protocol}

\author{
 {\rm Akshaya Kumar}\\
 Georgia Institute of Technology 
 \and
 {\rm Anna Raymaker}\\
 Georgia Institute of Technology 
 \and
 {\rm Michael A. Specter}\\
 Georgia Institute of Technology 
} 

\maketitle

\begin{abstract}
We conduct the first comprehensive security analysis of Tile, the second most popular crowd-sourced location-tracking service behind Apple's AirTags. 
We identify several exploitable vulnerabilities and design flaws, disproving many of the platform's claimed security and privacy guarantees: Tile's servers can persistently learn the location of all users and tags, unprivileged adversaries can track users through Bluetooth advertisements emitted by Tile's devices, and Tile's anti-theft mode is easily subverted. 

Despite its wide deployment---millions of users, devices, and purpose-built hardware tags---Tile provides no formal description of its protocol or threat model. 
Worse, Tile \emph{intentionally} weakens its antistalking features to support an antitheft use-case and relies on a novel ``accountability'' mechanism to punish those abusing the system to stalk victims. 

We examine Tile's accountability mechanism, a unique feature of independent interest; no other provider attempts to guarantee accountability. 
While an ideal accountability mechanism may disincentivize abuse in crowd-sourced location tracking protocols, we show that Tile's implementation is subvertible and introduces new exploitable vulnerabilities. 
We conclude with a discussion on the need for new, formal definitions of accountability in this setting.

\end{abstract}

\section{Introduction}

\begin{table*}[t]
    \centering
    \begin{tabular}{lcccc}
        \toprule
        \multirow{2}{*}{\textbf{Adversary}} & \multicolumn{4}{c}{\textbf{Attack Compromises}} \\ 
        \cmidrule(l){2-5}
            & Unlinkability & Anti-stalking alerts & Location privacy & Framing resistance  \\
            & (tag indistinguishability) & (tag detectability) & (location indistinguishability) & (framing resistance)  \\
        \midrule
        Passive RF\hfill      & \checkmark &            &            & \\
        Active RF\hfill       & \checkmark &            &            & \checkmark \\
        Malicious owner \hfill        &            & \checkmark &            &            \\
        Malicious tag \hfill       &            & \checkmark &            & \checkmark \\
        Platform server\hfill & \checkmark &            & \checkmark & \checkmark \\
        \bottomrule
    \end{tabular}
    
    \caption{A summary of attacks by adversary type. We show what kind of adversary is capable of executing what sort of attack; e.g., a passive RF adversary can break Tile's unlikability/tag indistinguishability property, allowing the attacker to link Bluetooth advertisements over arbitrary time periods to a device. The viability of these attacks may be dependent on the configuration of the tag. Our results are for the Tile Mate 2022. We define adversary types and security properties in \secref{subsec:sec-props-and-threat-models}.}
    \label{table:attacks}
\end{table*}

Tile is one of the most popular Bluetooth-based location-tracking services.
It has mature applications on both Android and iOS, and as of September 2021, Tile had sold over 40 million devices and had over 425,000 paying users~\cite{tile2021}.
The company has partnered with 19 third-party manufacturers---including Dell, Bose, and Fitbit---to embed its protocol into laptops, headphones, and smartwatches.
In June 2021, Tile devices were integrated into Amazon’s Sidewalk network, which claims coverage of 90\% of the US population~\cite{amazon2021_echo_tile_level,amazon2023_sidewalk_developers}. 
In November 2021, the family communication service Life360~\cite{Life360Homepage} acquired Tile for \$205 million, expanding Tile's network by 33 million smartphones \cite{life3602021}.

\textbf{In this paper, we present the first comprehensive security analysis of Tile's location tracking protocol.}
Our analysis reveals that Tile is vulnerable to several attacks that compromise the privacy and security of its users.
For example, we find that Tile's servers collect location information on millions of devices, effectively running a mass surveillance network (\secref{subsec:vulnsAgainstLocationPrivacy}).
Furthermore, we demonstrate that a third-party RF/network adversary can easily monitor a user's physical movements by passively observing the Bluetooth advertisements emitted by the user's Tile tracker (\secref{subsec:vulnsAgainstOwnerPrivacy}).
We also show that malicious users can exploit Tile trackers to stalk their victims (\secref{subsec:vulnsAgainstTagDetectability}),
and discover that Tile's novel Anti-Theft Mode and accountability mechanisms are susceptible to abuse and subversion (\secref{subsec:vulnsEnablingstalking} and \secref{subsec:vulnsAgainstAntitheft}).
We provide a detailed analysis and experimental verification for these vulnerabilities, showing that exploitation is within the reach of a motivated individual or malicious server.

Although Tile is one of the oldest Bluetooth-based crowd-sourced location tracking system, it competes with similar services from Apple, Samsung, and Google. 
Jointly with Tile, these vendors have sold hundreds of millions of devices.

As these cryptographic services, collectively known as \textit{\underline{O}ffline-\underline{F}inding} (OF) networks, become increasingly widespread and integrated into everyday devices, their potential for misuse is growing.
Bluetooth-based tracking tags have facilitated several criminal incidents, including theft, stalking, physical assault, and even murder~\cite{news1,news2,news3,news4,news5,news6,news7}. 
In these cases, a malicious user places their tag on a target (person or item) to monitor them via the OF network.

To prevent such abuse of OF networks, designers implemented ``anti-stalking'' features, allowing potential victims to detect rogue tags that may be traveling with them.
These include Apple's Item Safety Alerts, Samsung's Unknown Tag Alerts, and Tile's Scan and Secure feature. 

While these features help detect rogue tags, the questions remain: \textit{What recourse does a stalking victim have once they have discovered a malicious tag? How can abusers of OF networks be held accountable?} 
This is where Tile differentiates itself from other OF providers: it is the \textit{only} company that claims to offer accountability.

To contextualize Tile's accountability guarantees, we first describe its \textit{Anti-Theft} mode. 
In February 2023, Life360 CEO Chris Hulls claimed in an independent blog post~\cite{hulls_tile_approach_2024} that, ``\textit{an unintended consequence of these [antistalking] initiatives has been the neutering of the ability of Bluetooth locators to help recover stolen items after a theft}.''
The rationale was that a thief could run an antistalking scan, detect the tag, and discard it, eliminating its usefulness in recovering stolen items.
Soon after, Tile launched their \textit{Anti-Theft mode} feature, making Tile trackers undetectable by anti-stalking features.

To discourage abuse of the Anti-Theft mode for stalking, Tile implemented an accountability mechanism. Users of the Anti-Theft mode must verify their identity with a government-issued ID and selfie, and acknowledge that their information will be shared with law enforcement at Tile's discretion. 
They must also consent to a \$1 million fine if convicted of using Tile trackers for stalking.

While there are legal and practical questions surrounding the enforceability Tile's fine, there is academic interest in understanding the technical barriers and tradeoffs in implementing accountability in OF systems.
For example, an accountability mechanism must support dispute resolution, allowing a victim to prove they are not attempting to frame an honest owner, and avoid universally puncturing anti-surveillance and other privacy guarantees.  
Understanding Tile's design can provide a useful case study in real-world design of accountability in Offline Finding systems.

In addition to accountability, Tile makes other strong claims about their platform's security and privacy. 
Tile's privacy policy~\cite{TilePrivacyPolicy} claims that ``...[A]ny information transmitted across our network is anonymous. You are the only one with the ability to see your Tile location and your device location.'' 
However, Tile's privacy policy also links to Life360's general privacy policy, which appears to cover their entire suite of products and admits to a more expansive set of collection activities (including location data). 
\footnote{The Federal Trade Commission (FTC) has indicated via regulatory action that statements in privacy policies do not override a falsehood elsewhere~\cite{FTCsTwitterCase2022}.}

Despite their bold claims, there is little public information about the design of Tile's system. 
Tile's support page~\cite{TileHowItWorks} and privacy policy~\cite{TilePrivacyPolicy} provide only a vague description of its system and threat model.
Therefore, we reverse engineer Tile’s publicly available Android application to analyze the security of its OF network. 
For unknowable parts of Tile's private infrastructure, we make optimistic assumptions.
We show that the attacks we identify are exploitable even under these optimistic assumptions.  

\heading{Our Contributions}
We present the first comprehensive security analysis of Tile’s OF protocol. In particular, we:
\begin{itemize}[noitemsep, topsep=0pt]
    \item reverse-engineer and reconstruct Tile's protocols for tag registration, location reporting, Bluetooth advertisement generation, and its anti-stalking feature;
    \item identify and experimentally validate several vulnerabilities that allow adversaries with different capabilities to violate unlinkability, location privacy, and framing resistance, as summarized in~\Cref{table:attacks};
    \item analyze Tile’s novel Anti-Theft mode and accountability mechanism, showing that both can be circumvented and misused for stalking or framing attacks;
    \item compare Tile with other contemporary OF networks, such as Apple’s Find My and Google’s Find My Device, and show that it provides substantially weaker security.
\end{itemize}

\heading{Outline}
We begin~\secref{sec:background} with background on OF systems and their security, and state Tile’s claims of security.
We then summarize prior work in~\secref{sec:related-works}. 
In~\secref{sec:methodology}, we describe our reverse engineering methodology. 
In~\secref{sec:protocol-description}, we detail Tile’s protocol as discovered in
our methodology, covering registration, the OF protocol, custom cryptography, and a brief discussion of the parts we could not confirm. 
We present the attacks we identified against Tile's protocol in~\secref{sec:security-analysis}. 
We compare Tile's security and privacy guarantees with those of other major service providers like Apple and Samsung, and provide lessons learned and recommendations for incorporating accountability in OF systems in~\secref{sec:discussion}.
Finally, we conclude in~\secref{sec:conclusion}.
\section{Background}\label{sec:background}

We begin this section with an overview of a generic OF system, explaining the various entities involved, their interactions, and communication channels.
We then describe the threat model and security requirements commonly considered in the literature for both deployed and proposed OF systems, and with an overview of Tile's security claims. 

\subsection{Offline finding systems}\label{subsec:of-overview}
We briefly sketch the devices and steps used by OF systems.  
We begin by broadly describing Bluetooth Low Energy (BLE) communication that forms the core of OF protocols.

\heading{Bluetooth Low Energy (BLE)}
BLE communication occurs through advertisements or full connections.
Advertisements are connectionless and used for device discovery and connection initiation. 
Advertising packets contain devices' identity, capabilities, and services.
These packets are received by nearby devices scanning for BLE advertisements.
Connections allow bidirectional communication, and are established through the \underline{G}eneric \underline{Att}ribute Profile (GATT) protocol~\cite{bluetoothGATT}.

\begin{figure}[t]
    \centering
\begin{tikzpicture}

\tikzstyle{device} = [draw, minimum size=1.5cm]
\tikzstyle{service} = [cloud, draw, minimum size=1.5cm, aspect=2]
\tikzstyle{arrow} = [->, thick]

\node[minimum size=1cm] (owner1) {\includegraphics[width=1.1cm]{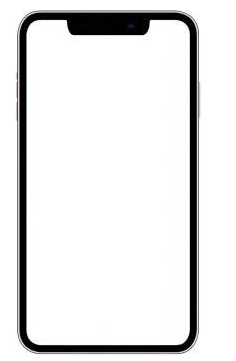}};
\node[below=1mm of owner1] (ownertext) {\small{owner device}};
\node[minimum size=1cm, below=2cm of owner1] (tile_connected) {\includegraphics[width=1cm]{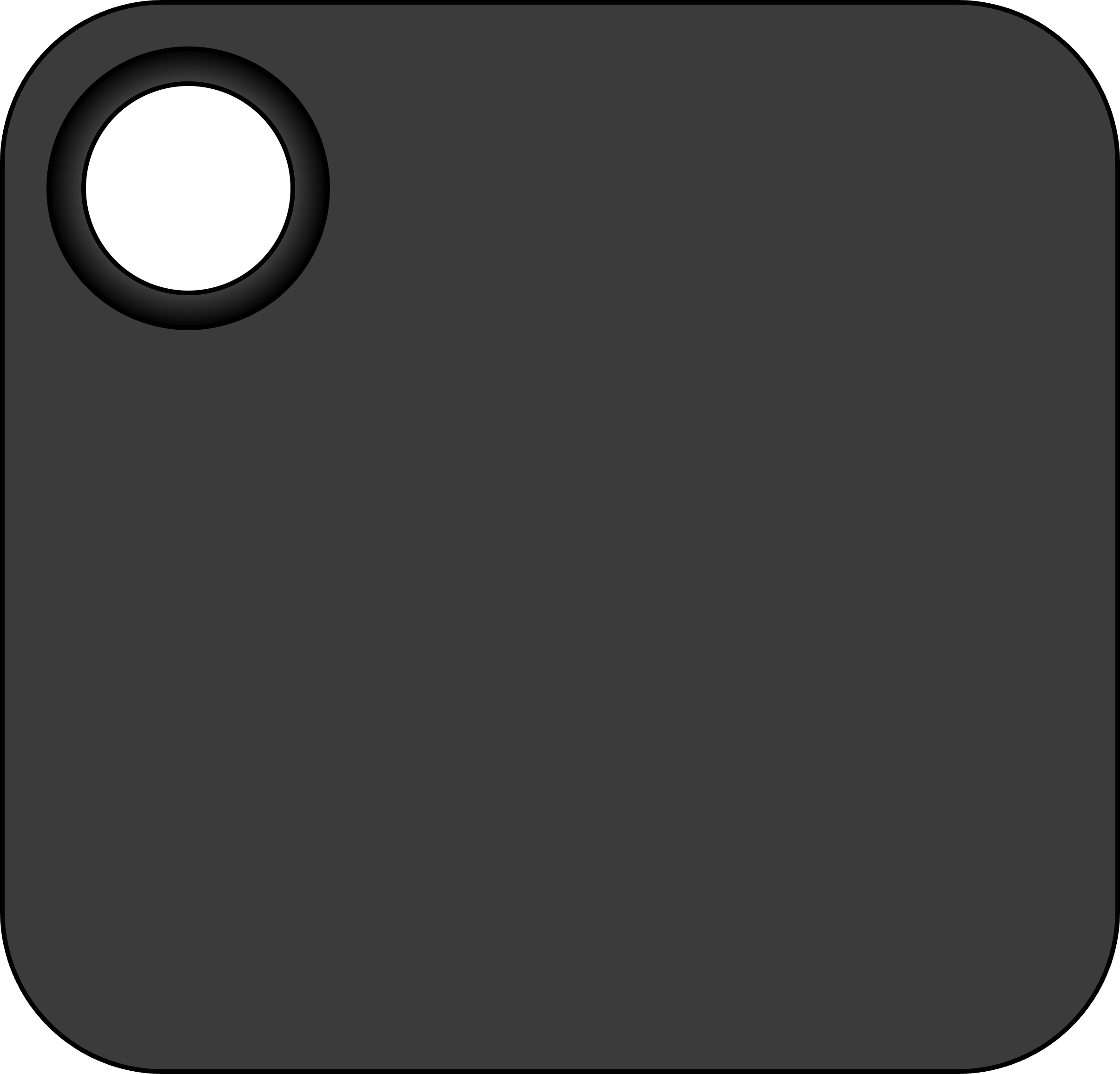}};
\node[align=center, below=1mm of tile_connected] {\small{tracker tag}\\ \small{(connected)}};
\draw[arrow] (ownertext) -- node[left] {\small{BLE connection}} 
                            node[right] {\includegraphics[width=0.3cm]{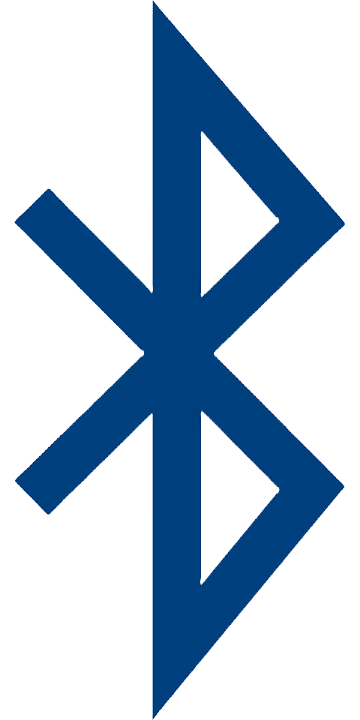}}(tile_connected);
\draw[arrow] (tile_connected) -- (ownertext);

\node[minimum size=1cm, left=0.25cm of tile_connected] (tile_lost) {\includegraphics[width=1cm]{tag.png}};
\node[align=center, below=1mm of tile_lost] {\small tracker tag\\\small(lost)};
\node[minimum size=1.5cm, left=3cm of tile_lost] (finders) {\includegraphics[width=1.3cm]{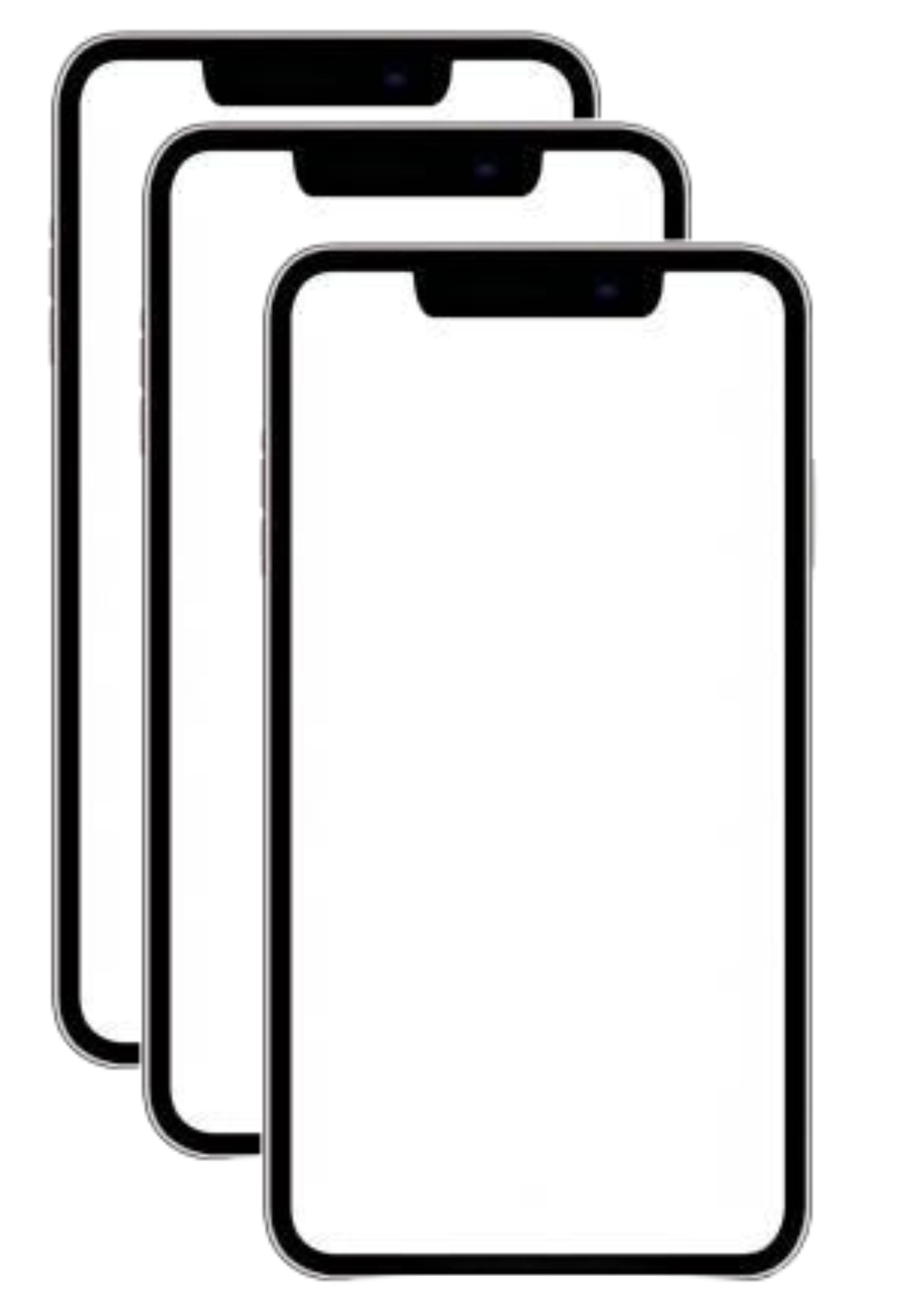}};
\node[below=1mm of finders] (textfd) {\small{finder devices}};

\draw[arrow] (tile_lost) -- node[above] {\small{BLE advertisements}} 
                            node[below] {\includegraphics[width=0.3cm]{bluetooth.png}}
                            (finders);

\node[minimum size=1.5cm, above=1.75cm of finders] (service_provider) {\includegraphics[width=1.5cm]{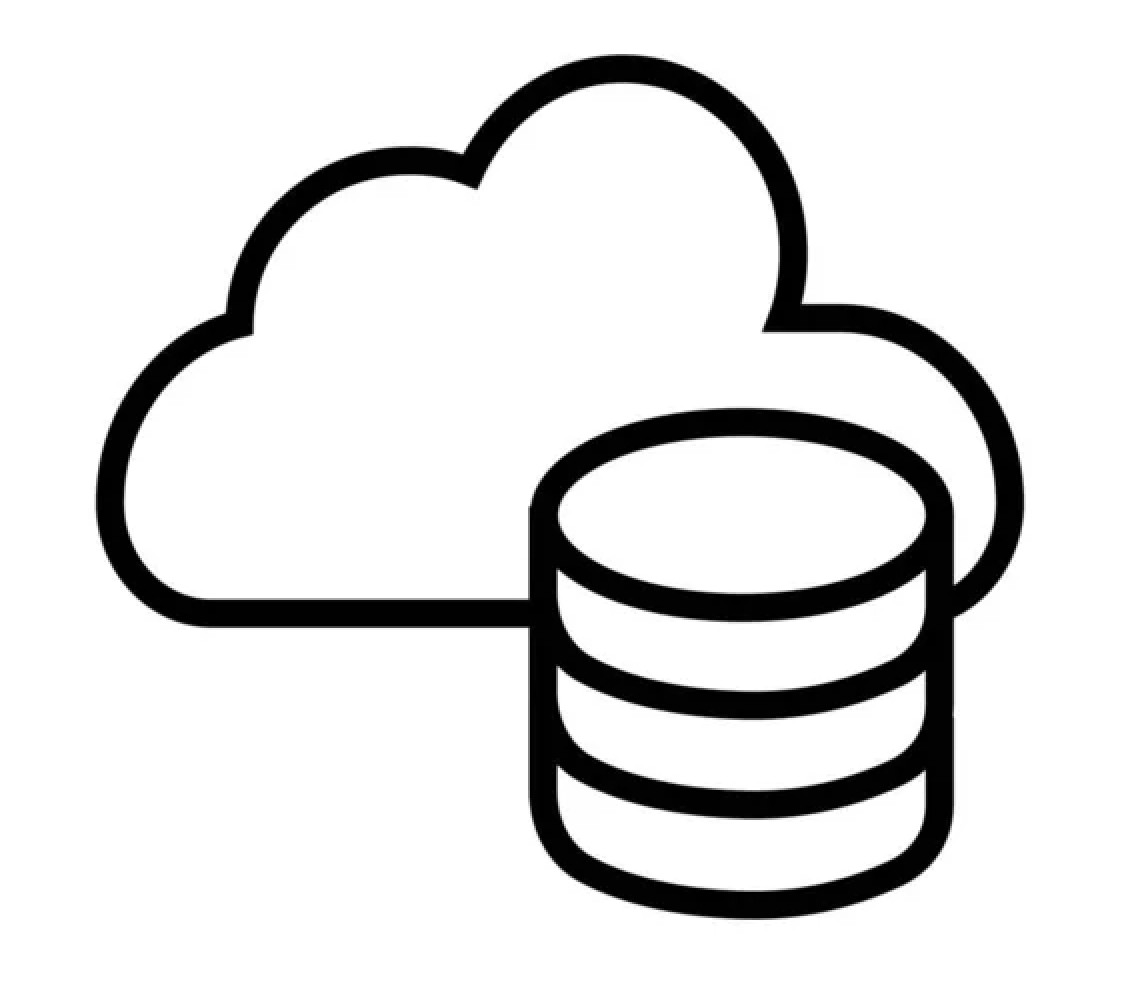}};
\node[below=1mm of service_provider] (sptext) {\small{service provider}};

\draw[arrow] (finders) -- node[right] {\small{location reports}} 
                          node[left] {\includegraphics[width=0.6cm]{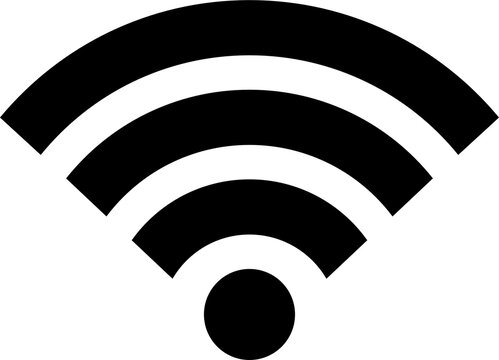}}(sptext);

\draw[arrow] (service_provider) -- node[above] {\small{download location reports}} 
                                   node[below] {\includegraphics[width=0.6cm]{wifi.jpg}}
                                   (owner1);

\end{tikzpicture}
\caption{The components involved in an OF protocol.}
\label{fig:OF-overview} 
\end{figure}

\heading{An overview of OF protocols}
We summarize the components involved in an OF protocol and their interactions in~\Cref{fig:OF-overview}.
First, a user buys a \textit{tracker tag}, a small battery-powered device with low memory and computing power that has BLE capabilities but no access to the internet.
Then, they pair the tracker tag with \textit{owner devices} like smartphones or tablets that are more powerful than tags; they have GPS and internet capabilities.
A user’s owner devices share the same account.
Owner devices and associated tags communicate with each other over a BLE connection.
A tag is considered ``lost'' when it loses Bluetooth connectivity with its owner device (otherwise, it is considered ``connected'').
Similarly, an owner device is considered lost if it disconnects from the internet.
A lost tag/device sends out BLE advertisements containing an identifier to be discovered by finder devices.
\textit{Finder devices} are bystander devices on the service provider’s network that have GPS and internet capabilities and scan for BLE advertisements to discover lost tags/devices.

Once a lost tag/device is detected, the finder records its own location and sends it to the service provider's servers along with the lost tag/device's identifier.
The \textit{service provider} maintains a database of location reports and identifiers submitted by finder devices.
Finally, an owner device can retrieve the locations of their lost tags/devices by querying the service provider for their tags' identifiers.

\subsection{Security properties and threat models}\label{subsec:sec-props-and-threat-models}
The security of OF networks is an emerging area of research, with several recent papers studying existing networks and proposing new constructions.
In this section, we present security properties and corresponding threat models, adopting definitions from prior work, and define \emph{framing resistance}, a new definition that is a necessary precursor to accountability in OF systems. 

\heading{Tag indistinguishability against passive and active RF adversaries}
If a tag consistently broadcasts a unique identifier (e.g., a static MAC address), then an adversary can easily link the identifier to its owner and track their movements.

A \textit{passive RF adversary} is an attacker equipped with a Radio Frequency (RF) detector, allowing inspection of Bluetooth emissions from nearby tracker tags.
The passive RF adversary is also referred to as the ``tracking adversary''~\cite{299830}, and could use information learned from Bluetooth advertisements to ``track'' the corresponding owner.
In the worst case, an RF adversary could set up a network of RF sniffing devices to listen for BLE advertisements broadcast by several thousand trackers across a large area, deanonymize corresponding owners, and monitor their movements.
An \textit{active RF adversary} has the capabilities of the passive RF adversary, but can additionally emit arbitrary Bluetooth packets. 

Manufacturers protect the privacy of owner devices by designing their tags to emit periodically changing identifiers such that any two identifiers broadcast by a tag are cryptographically indistinguishable from identifiers broadcast by two different tags. 
This property is formally referred to as \textit{tag indistinguishability}~\cite{Mayberry2023BlindM,299830}. 

\heading{Tag detectability against malicious owners and tags}
A \textit{malicious owner} is an adversary that surreptitiously attaches its own tag to a victim and leverages the OF network to monitor the victim's movements.
This type of adversary is also known as the ``stalking adversary''~\cite{299830}. 
One can think of a malicious owner as an honest-but-curious adversary who does not deviate from the protocol specification but tries to learn information it is not privy to.  
This cannot be fully prevented, as it is a natural consequence of the ability to track tags, but OF networks should be designed to be resistant to such abuses.

To prevent network abuse for stalking, manufacturers implement anti-stalking features that guarantee \textit{tag detectability}~\cite{299830}, allowing a user's device to determine if identifiers recorded at different locations and times originated from the same tag.
Apple and Google's OF systems disable the periodic rotation of a tag's identifier once it is separated from the owner's device, forcing it to advertise static identifiers.
Intuitively, this weakens the tag indistinguishability property, but 
allows the victim's device to alert the user if the same (unknown) identifier is seen over a suspiciously long time and distance.

A \emph{malicious tag} adversary builds counterfeit tags to bypass stalking protections (e.g., by continuing to periodically rotate tag identifiers even in the lost mode). 
These counterfeit tags otherwise mimic the functionality of manufacturer-produced tags and can be tracked using the provider’s network, but remain undetected by anti-stalking features. 
Ideally, an OF system should offer tag detectability against both malicious owners and malicious tags.

\heading{Location indistinguishability against the platform server}
To prevent the server from learning the locations of the finders and owners based on the reports submitted, an OF system should provide \textit{location indistinguishability}~\cite{Mayberry2023BlindM,299830}.
This property guarantees that the server cannot learn location information from reports submitted by finder devices. 
Providers like Apple and Google achieve location indistinguishability by end-to-end encrypting location information using a public key embedded in BLE advertisements emitted by a tag.

\heading{Accountability and framing resistance}
While stalking resistance allows for detection of a stalker's tag, an \textit{accountability mechanism} allows the victim, platform, or an external authority to disincentivize misuse of the platform by punishing bad behavior.
This can be achieved by cryptographically tying the identity of owners to the protocol so that, in the event of misuse, a victim collaborating with appropriate authorities can pierce the tag indistinguishability property to punish the perpetrator. 

A core challenge is ensuring that an accountability mechanism does not allow an attacker to provide convincing proof that an innocent user owns a tag --- 
we call this property \emph{framing resistance}.

\subsection{Tile's claims of security}\label{sec:tiles_claims_of_security}
Although there is no public, formal description of their protocol or threat model, Tile makes several claims about their system’s security properties via their privacy policy~\cite{TilePrivacyPolicy}, support pages~\cite{TileScanSecureOverview,TileAntiTheftMode,TileNetwork}, blog~\cite{TileAntiTheftModeBlog,TileAntiTheftBlog}, and FAQ~\cite{TileScanAndSecureFAQ}.

\heading{Tag indistinguishability}
Tile claims that ``any information transmitted across our network is anonymous''~\cite{TilePrivacyPolicy} and ``Location updates are automatic and anonymous, so it’s completely secure''~\cite{TileNetwork}.

While explaining the results of Scan and Secure~\cite{TileScanAndSecureFAQ}, Tile claims that ``The Reference ID [returned by Scan and Secure] is a unique and encrypted identifier of the Tile or Tile-enabled device that intermittently changes to ensure the privacy of the Tile owner. 
For example, if another person running the Scan and Secure feature detected a Tile that you own, they would not be able to identify you as the Tile owner.''

\heading{Location indistinguishability}
In its privacy policy~\cite{TilePrivacyPolicy}, Tile states that ``You are the only one with the ability to see your Tile location and your device location.'' 
Furthermore, they claim that, ``You won't be able to see where other people's Tiles are, and they won’t be able to see yours''~\cite{TileNetwork}.

\heading{Tag detectability}
Tile guarantees that their Scan and Secure feature allows users to ``detect nearby Tiles and Tile-enabled devices'' that may be traveling with them. 

\heading{Tag (un)detectability in Anti-Theft mode}
Tile promises that once a user activates the Anti-Theft mode, all their Tiles ``will no longer be discoverable by our Scan and Secure feature and certain other 3rd party scanning tools''~\cite{TileAntiTheftMode}.

\heading{Accountability}
To prevent the misuse of the Anti-Theft mode for stalking, Tile claims to have implemented ``strict safety measures.''
This includes requiring a user to verify their identity using a government ID and a live photo before activating Anti-Theft mode.
Tile also states that they will impose a ``\$1 million fine for any individual convicted in a court of law for using Tile devices to illegally track any individual without their knowledge or consent.'’
It is unclear how enforceable the above agreement is or how such a proposal would work in a region without the rule of law.

Somewhat alarmingly, Tile has made inconsistent statements regarding when it will share information with law enforcement.
On the one hand, Tile specifies in the FAQ page~\cite{TileScanAndSecureFAQ} that they will ``work with law enforcement through a properly issued court order to identify the owner of a suspicious Tile using the Reference ID.''
On the other hand, they require users of the Anti-Theft mode to agree that their ``personal information can and will be shared with law enforcement at our discretion, even without a subpoena.''

Tile asserts that their system can generate legally admissible proof that a device used for stalking is owned by a particular user, without detailing the validity of this proof.
This capability is directly presented as a solution to the weaknesses introduced by the Anti-theft mode.
While Tile's policy avoids explicitly using the term ``accountability,'' this assertion is specifically framed around Tile's protocol for identifying and punishing perpetrators of malicious tracking.

\section{Related work}\label{sec:related-works}
Previous studies have examined Tile's protocol only in passing, within the broader context of Bluetooth-based tracking technologies. 
None have analyzed its overall threat model, protocol, potential for surveillance, or accountability mechanism, anti-stalking, and Anti-theft features.
We are the first to analyze Tile's protocol exhaustively and find the vulnerabilities presented in~\secref{sec:security-analysis}.

\heading{Prior work on Tile}
Weller et al.~\cite{Weller_2020} performed the first security and privacy analysis of several Bluetooth trackers, including Tile. 
They reported that the app sends metadata---including the currently used Wi-Fi name and Wi-Fi MAC address---to the server.
Garg et al.~\cite{10.1145/3448300.3467821} outlined a set of desirable security guarantees for crowd-sourced location tracking systems and analyzed whether various services, including Tile, met these guarantees, but provided limited analysis on each. 
Various works~\cite{PACE2023301559,d204n6_android_tile_2019,d204n6_ios_tile_2020,gazeau2020catch} have analyzed the forensic information stored by Tile on a user's phone. 
These studies revealed that an entity with temporary physical access to a user's Tile-enabled phone can recover (timestamped) location information about the user, their Tiles, and even other users for the past 30 days.

Heinrich et al.~\cite{heinrich2024popets} and Turk et al.~\cite{Turk2023CantKT} examined Tile as part of a broader study on the effectiveness of anti-stalking mechanisms employed by various OF service providers. They concluded that the guarantees provided by Tile’s Scan and Secure feature were inadequate in their threat models.

\heading{Security analysis of other OF networks}
Several works have analyzed the security of widely deployed OF protocols, including Apple and Samsung's implementations.

Heinrich et al.~\cite{DBLP:journals/corr/abs-2103-02282} reverse-engineered Apple's FindMy protocol and provided its first technical specification.
Their work revealed design and implementation flaws that could potentially result in location correlation attacks and unauthorized retrieval of location histories.
They also implemented the OpenHaystack tool~\cite{10.1145/3448300.3468251} that helps create and use custom FindMy devices.
Mayberry et al.~\cite{10.1145/3463676.3485616} and Heinrich et al. ~\cite{10.1145/3507657.3528546} analyzed the security of Apple's Item Safety Alerts and presented ways to circumvent these alerts using custom tags.
Bräunlein showed the use of the FindMy network for covert communication~\cite{websiteSendMy}.
Yu at al.~\cite{yu2022privacy} and Liu et al.~\cite{LiuUsenix2025} bfound several vulnerabilities in their analysis of Samsung's OF protocol which focused on the unlinkability of tags and privacy of location reports.
The security of both Apple's and Samsung's OF protocols has also been analyzed in a broader threat model~\cite{alamleh2024securing}.
Google's protocol was recently reverse-engineered by Botteger et al.~\cite{okayGoogle}, revealing vulnerabilities that allow a malicious owner to evade detection and compromise privacy. 
All these vulnerabilities resulted from bad design, just as we will see is the case with Tile. 

\heading{Proposed (cryptographic) solutions for OF systems}
Mayberry et al.~\cite{Mayberry2023BlindM} formalized security notions for a crowd-sourced location tracking system.
They propose a modification of the cryptographic protocol underlying Apple's FindMy network that achieves their security notions.
In particular, their protocol leverages partially blind signatures to guarantee that that attackers cannot add non-server authorized malicious tags to the network.

Eldridge et al.~\cite{299830} formalize strong security definitions for privacy-preserving abuse-resistant location tracking protocols that guarantee robust stalker detection whilst maintaining user privacy.
They also construct a new OF protocol using secret-sharing and lattice-based algorithms as building blocks.
Engineers at Google and Apple have created Detecting Unwanted Location Trackers (DULT)~\cite{detecting-unwanted-location-trackers-01}, a draft IETF specification.
David et al.~\cite{11023275} present advanced security and privacy notions for OF protocols and present XDHIES, a cryptographic tool that enables achieving these properties.

\heading{Anti-stalking studies}
Several studies have analyzed anti-stalking countermeasures implemented by service providers. 
Heinrich et al.~\cite{10.1145/3507657.3528546} analyzed Apple's anti-stalking mechanism and found that, while Apple's solution protects iOS users, their anti-stalking application for Android was insufficient.
They developed Airguard, an Android application that better detects the presence of rogue AirTags around Android users. 
Airguard also detects rogue tags from other manufacturers, including Samsung and Tile.
M{\"u}ller et al.~\cite{muller2023homescout} analyzed the potential for misuse of various trackers and developed a similar application. 

In recent work~\cite{heinrich2024popets}, Heinrich et al. measured the misuse of Bluetooth trackers, particularly Apple, Tile, Samsung, and Chipolo, for unwanted tracking and stalking. 
The study used data from the AirGuard app and a large user survey.
It reported that over 40\% of stalking victims had been subjected to location tracking by tracker tags.
Turk et al.~\cite{Turk2023CantKT} also investigated the effectiveness and limitations of several service providers' antistalking mechanisms. 
They highlighted that in many cases, victims do not use these features, even if they suspect that they are being stalked.

\heading{Tracking BLE devices}
Celosia et al.'s analysis of the BLE advertising mechanism~\cite{10.1145/3360774.3360777} showed that advertisements may contain information that uniquely identifies a Bluetooth device, despite MAC address randomization. 
They also note that many devices continue to broadcast static addresses, even though identification and subsequent tracking of Bluetooth devices emitting static MAC addresses is a known issue.

\section{Methodology}\label{sec:methodology}
In order to gain a thorough understanding of Tile's infrastructure, we began by decompiling the most recent version (2.125.0) of their Android application as found on the Google Play Store.
We used a Google Pixel 3XL, jailbroken using Magisk~\cite{topjohnwu_magisk}, as our test device. 
For our experiments, we used the Tile Mate 2022 running the pre-activation firmware version 48.04.16.0 and post-activation firmware version 48.04.28.0.

We then studied the decompiled code to learn various aspects of Tile’s protocol, including the registration process, generation of cryptographic secrets, location reporting and retrieval, Scan and Secure, and Anti-Theft protocols.
Additionally, we performed dynamic analysis to inspect BLE messages exchanged between the Tile and other devices, and network traffic analysis to inspect HTTP messages sent between tags, devices, and the server.
Our experimental methodology for verifying the vulnerabilities we describe in~\secref{sec:security-analysis} overlapped with the dynamic analysis we performed in the reverse engineering process.
We describe any differences in experimental methodology in~\secref{sec:security-analysis}.

\heading{Static analysis}
We performed static analysis on Tile's Android application to study the closed-source implementation of Tile's OF protocol.
In particular, we used the JADX~\cite{jadx} decompiler to convert the Tile APK into Java-like source code.
We then analyzed the obtained source code to piece together Tile's OF protocol.

\heading{Android Bluetooth log analysis}
By enabling the Bluetooth HCI snoop log option, Android records Bluetooth communication between the mobile device and peripherals.
This allowed us to examine data exchanged between a phone and a tag over BLE.

\heading{Network analysis}
We analyzed the messages exchanged between a Tile-enabled Android client with other devices/parties by monitoring network traffic.
To monitor HTTPS traffic between our test device and the server, we used the BurpSuite tool to set up a proxy to man-in-the-middle and decrypt all incoming and outgoing traffic from our device.

\heading{Limitations of our study}
Our study focuses exclusively on Tile's most common tracker, the Tile Mate 2022. 
Tile manufactures several other models and also collaborates with third-party manufacturers.
Our findings might not apply universally across all devices.
Furthermore, our analysis does not include access to Tile's server backend, limiting our understanding of server-side processes.
We make best-case assumptions regarding those operations in our analysis of their protocol.

\section{Tile's offline finding protocol}\label{sec:protocol-description}
In this section, we outline the OF protocol executed between a Tile tracker (tag), a phone that is paired and connected to the tracker (owner), the Tile server(s), and a network of bystander devices using the Tile application (finders).
We break the protocol down into multiple phases --- owner/finder registration (\secref{subsec:owner-finder-registartion}), tag activation (\secref{subsec:tag-activation}), tag-owner interactions (\secref{subsec:tag-owner-interactions}), tag-finder-server interactions (\secref{subsec:tag-finder-server-interactions}), owner-server interactions (\secref{subsec:owner-server-interactions}), Scan and Secure (\secref{subsec:scan-and-secure}), the Anti-Theft mode (\secref{subsec:antitheft-mode}), tag transfers and sharing (\secref{subsec:tag-transfers-and-sharing}), Community Information (\secref{subsec:community-info}), and account deletion (\secref{subsec:account-deletion}).

We begin by providing a brief overview of the protocol.
The first step for the owner, the tag, and the finder is registration.
For the owner and the finder, this involves creating an account with Tile using an email address and a password over an active network connection.
The tag is activated by registering it with an owner device.
The owner acts as a proxy between the tag and the server during this step.
An activated tag transmits BLE advertisements containing a unique and rotating identifier.
These advertisements are picked up by finder devices that extract the identifier from the BLE advertisements and store them in a local database.
Once the finder connects to the internet, it uploads the recorded identifiers along with its current location to the Tile server.
Finally, the owner of a tag can query the server for location information about its tag.
We explain each of these steps in detail below.

\subsection{Owner/Finder registration}\label{subsec:owner-finder-registartion}
We now describe the registration process for owner and finder devices. In the Tile ecosystem, these devices are smartphones or tablets.

To register, an owner first downloads the Tile Android application and creates an account using an email address and password.
During this process, the owner's Android device generates a static 16-byte $\variable{client\_uuid}$ that is unique to the particular device. 
Then the device sends a $\mathsf{POST}$ request to the Tile API server at production.tile-api.com/api/v1/users, containing $\variable{client\_uuid}$ and the owner's email ID and password.
The user is sent a 6-digit verification code on the email ID they provided.
Notably, the user isn't required to prove ownership of the email; this step is skippable. 
The response from the server includes a 16-byte $\variable{user\_uuid}$ assigned by the server and contains a status field whose value is set to ``ACTIVATED''.
The $\variable{user\_uuid}$ is used to identify the owner in future communications with the server.
Next, the device issues another $\mathsf{POST}$ request at /api/v1/tiles/generate\_tileUUID, containing the $\variable{tile\_uuid}$, the $\variable{user\_uuid}$, and $\variable{tile\_type}$ where $\variable{tile\_uuid}$ is set to $\variable{client\_uuid}$, $\variable{user\_uuid}$ is the value previously returned by the server, and $\variable{tile\_type}$ is set to ``PHONE''. 
The server's response overwrites $\variable{tile\_uuid}$ with a fresh 16-byte value prefixed with the string ``p!'' for device-type identification.
This completes registration for the owner. 

\subsection{Tag activation}\label{subsec:tag-activation}
\begin{figure}[t] 
\centering
\small
\begin{tikzpicture}
\small{
\tikzstyle{arrow} = [->, thick]
\tikzstyle{block} = [rectangle, draw, fill=white, text centered, minimum height=2em, minimum width=2em]

\node (tile_owner) at (0,0) {
    \includegraphics[width=2.5cm]{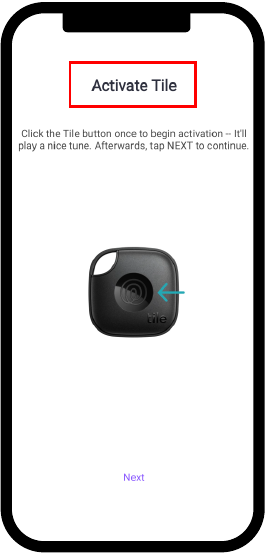}
};

\node (tile_tag) at (6,1.2) {
    \includegraphics[width=1.5cm]{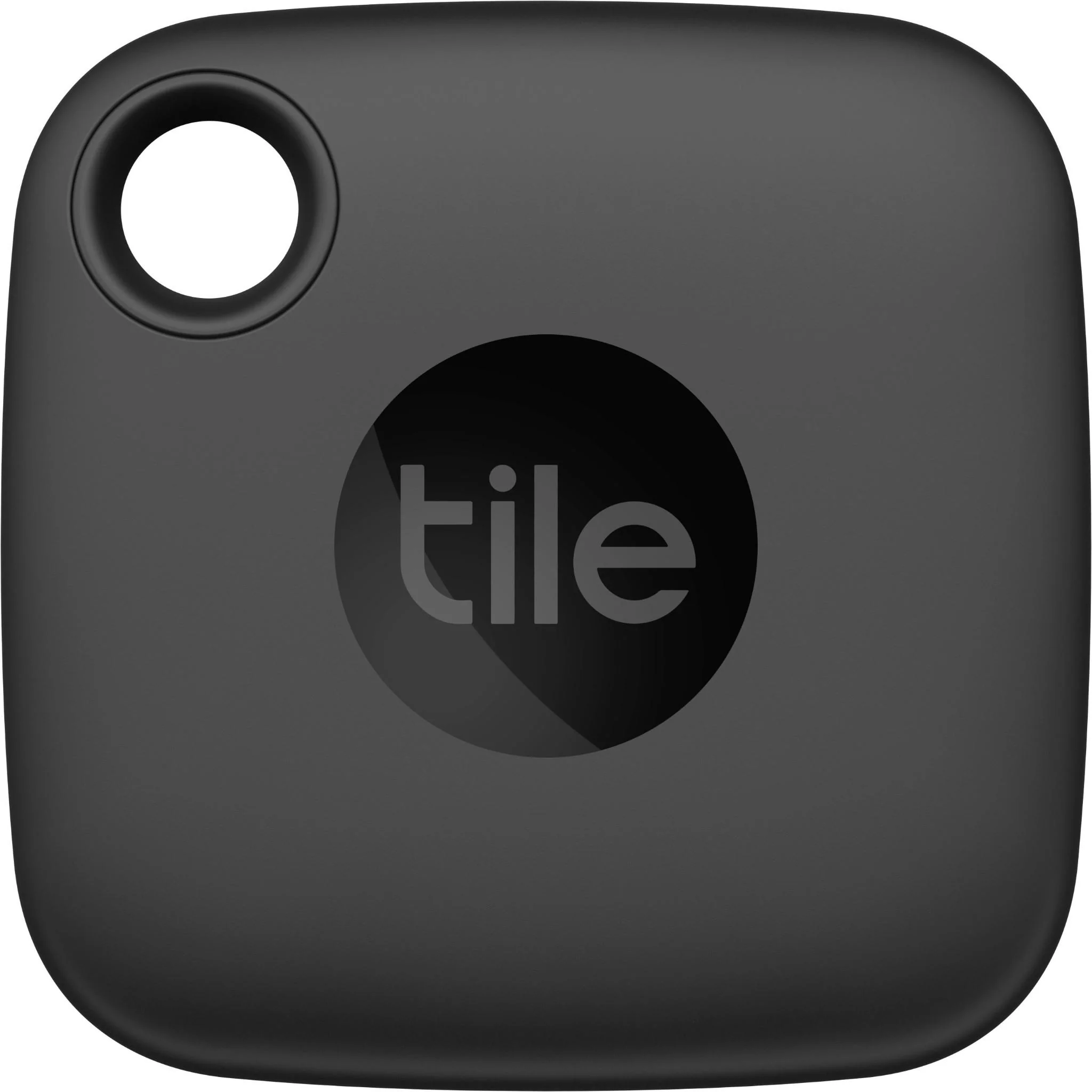}
};

\node (tile_server) at (6,-1.5) {
    \includegraphics[width=2cm]{service-provider.png}
};

\draw[arrow] (1.5, 2) -- (4.8, 2) node[midway, above] {1. Request TDI};
\draw[arrow] (4.8, 1.5) -- (1.5, 1.5) node[midway, above] {2. TDI};
\draw[arrow] (1.5, 1) -- (4.8, 1) node[midway, above] {3. $\variable{randA}$};
\draw[arrow] (4.8, 0.5) -- (1.5, 0.5) node[midway, above] {4. $\variable{randT}$, $\variable{sresT}$};

\draw[arrow] (1.5, -1.3) -- (4.8, -1.3) node[align=center, midway, above] {5. TDI, $\variable{randA}$, \\ $\variable{randT}$, $\variable{sresT}$};

\draw[arrow] (4.8, -1.8) -- (1.5, -1.8) node[midway, above] {6. $\variable{authKey}$};

}
\end{tikzpicture}
\caption{An overview of the various steps involved in Tile tracker activation.}
\label{fig:tile-activation}
\end{figure}

We now describe the process by which a Tile tracker is activated and associated with an owner device. 
We summarize the interactions between the owner device, the tracker tag, and the server during the tag activation phase in~\Cref{fig:tile-activation}.
We describe each step below.

\heading{Tile Device Information (TDI) [Steps 1,2 in \Cref{fig:tile-activation}]} Before a Tile tracker is activated, it broadcasts advertisements containing the pre-activation UUID `FEEC'.
The tracker owner initiates a Bluetooth scan through the Tile Android app to activate it.
The results of this scan are filtered using the `FEEC' UUID. 
The owner requests the $\variable{tileId}$, $\variable{model}$, $\variable{firmware}$, and $\variable{hardware\_version}$ from the tag over the custom Tile Device Information (TDI) service (UUID 180A). 
The corresponding characteristic UUIDs for these variables are defined in~\Cref{table:1} in~\Cref{sec:char-uuids}.

\heading{Tag authentication [Steps 3,4 in \Cref{fig:tile-activation}]}
During the activation process, the owner establishes a shared secret called the $\variable{authKey}$ between itself, the server, and the Tile tag.
To initiate this process, the owner's phone connects with the GATT server on the tag.
After the connection is established, the tracker exposes its `FEED' service to the owner, and the owner and the tracker execute a challenge-response protocol for authenticating the tag to the server.

The owner samples a 14-byte uniformly random value called $\variable{randA}$ using Java's SecureRandom random number generator and sends it to the tag.
Subsequently, the Tile tag generates a 10-byte random value $\variable{randT}$.
The tag then computes a 4-byte value $\variable{sresT}$ by applying the $\variable{HMAC\mbox{-}SHA256}$ hash function to $\variable{randA}$, $\variable{randT}$, and $\variable{tileId}$ using a 16-byte interim key $\variable{interimAuthKey}$ that is provisioned to a vendor by Tile and stored on device at manufacture time.
Essentially, during this step, the tag proves knowledge of the $\variable{interimAuthKey}$ to the server (using the owner device as a proxy) to authenticate itself as a legitimate device. 

\heading{Secret key establishment [Steps 5,6 in \Cref{fig:tile-activation}]}
The tag derives the permanent secret key---$\variable{authKey}$---by applying the $\variable{HMAC\mbox{-}SHA256}$ hash function to the $\variable{interimAuthKey}$ and the $\variable{sresT}$ value obtained above.
The $\variable{authKey}$ is the shared secret between the owner device and the tag.
However, the owner device does not know the $\variable{interimAuthKey}$ and cannot derive $\variable{authKey}$ itself. 
Instead, the owner device creates an HTTPS POST request to the server containing the information obtained during TDI as well as $\variable{randA}$, $\variable{randT}$, and $\variable{sresT}$.
We describe the structure of the request body in~\Cref{sec:http-request-bodies}.

The server first checks that $\variable{randA}$, $\variable{randT}$, and $\variable{sresT}$ form an unused and valid authentication triplet for the queried Tile tracker by checking the relation between them.
If the check fails, the server rejects the query.
Otherwise, the server derives the 16-byte $\variable{authKey}$ analogously to the tag, and returns the $\variable{authKey}$ to the owner.
This step completes the establishment of the $\variable{authKey}$ between the tag, the owner, and the server.
It is important to note that the server knows the $\variable{authKey}$ for every tag.
The $\variable{authKey}$ is subsequently used to produce pseudonymous identifiers for the tag to broadcast.

\heading{Generation of pseudonymous identifiers}
The owner device, tag, and server use the $\variable{authKey}$ to derive 8640 identifiers, called $\variable{privateIds}$, for the tag.
The $\variable{privateIds}$ are generated as follows where $\variable{tileId}$ is the unique identifier for the tracker tag, $\variable{identityBytes}$ is the little Endian byte encoding of the string ``identity'', and $\variable{ctr}$ is a counter initialized to 0 and incremented after each invocation of $\variable{HMAC\mbox{-}SHA256}$.
The concatenation $\variable{tileId}\concat\variable{identityBytes}$ is padded to 32 bytes.
The output of $\variable{HMAC\mbox{-}SHA256}$ is truncated to the first 64 bits.

\begin{center}
  $\variable{seed} \gets \variable{HMAC\mbox{-}SHA256}(\variable{authKey}, \variable{tileId}\concat\variable{identityBytes})$\\
  $\variable{privateId} \gets \variable{HMAC\mbox{-}SHA256}(\variable{seed}, \variable{ctr})[0:64]$
\end{center}

The tag continuously emits BLE advertisements containing a $\variable{privateId}$, starting at the value corresponding to the counter 0, and rotating to the next value every 15 minutes. 
With 8640 unique values, this rotation cycle repeats every 90 days.

\subsection{Tag-Owner interactions}\label{subsec:tag-owner-interactions}
\begin{figure}[t]
\centering
\small
\begin{tikzpicture}
\small{
\tikzstyle{arrow} = [->, thick]
\tikzstyle{block} = [rectangle, draw, fill=white, text centered, minimum height=2em, minimum width=2em]

\node (tile_tag) at (5,0) {
    \includegraphics[width=1.5cm]{tile.png}
};

\node (tile_owner) at (0,0) {
    \includegraphics[width=2.5cm]{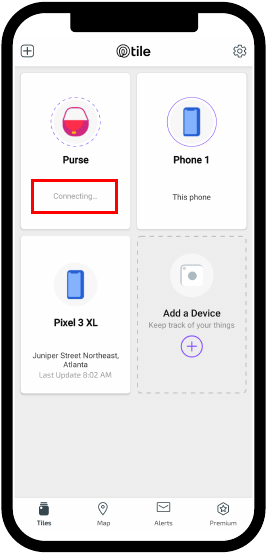}
};

\draw[arrow] (1.4, 0.3) -- (4, 0.3) node[midway, above] {\small{1. $\variable{randA}$}};
\draw[arrow] (4, -0.3) -- (1.4, -0.3) node[midway, above] {\small{2. $\variable{randT}$, $\variable{sresT}$}};
}

\node[align=center] (equation) at (4.3, -2) {
     $\variable{|randA}| = 14; |\variable{randT}| = 10; |\variable{sresT}| = 4$\\
     $\variable{|randA\mbox{-}pad}| = |\variable{randA\mbox{-}pad}| = 16$\\
     $\variable{seed} \gets  \variable{randA\mbox{-}pad} \concat \variable{randT\mbox{-}pad}$\\
   {$\variable{sresT} \gets \variable{HMAC\mbox{-}SHA256}(\variable{authKey},\variable{seed})[32:64]$}
};
\end{tikzpicture}
\caption{The authentication protocol used by the Tile tracker to authenticate itself to the owner device.}
\label{fig:tile-auth}
\end{figure}

In this subsection, we discuss the interactions between the owner device, the tag, and the server after the tag has been activated.
Most of these interactions facilitate ancillary features of the tag, such as ringing the tag and using the tag to (reverse) ring the owner.
In particular, these steps---with the exception of location reporting, which we describe in this section---are not relevant to the OF protocol.
We include them in \secref{sec:near-mode-tag-owner-interactions} for completeness.

\heading{Establishing a connected channel: Tag authentication}
Following tag activation, all tag-owner communications are transmitted over a connected channel.
In order to establish a connection, the owner and the tracker authenticate each other using Tile's custom challenge-response protocol.
We describe owner authentication in~\Cref{{sec:near-mode-tag-owner-interactions}}.
Tag authentication follows the protocol described in~\Cref{fig:tile-auth}.
This protocol is similar to Steps 3 and 4 in~\Cref{fig:tile-activation}.
The only difference is that the tag uses the $\variable{authKey}$ instead of the $\variable{interimAuthKey}$ to derive $\variable{sresT}$ from $\variable{randA}$ and $\variable{randT}$ as follows.
Bytes 4-7 of the $\variable{HMAC\mbox{-}SHA256}$ output are assigned to $\variable{sresT}$.

\begin{center}
    $\variable{seed} \gets  \variable{randA\mbox{-}pad} \concat \variable{randT\mbox{-}pad}$\\
   {$\variable{sresT} \gets\variable{HMAC\mbox{-}SHA256}(\variable{authKey},\variable{seed})[32:64]$}
\end{center}

Here $\variable{randA\mbox{-}pad}$ and $\variable{randT\mbox{-}pad}$ are obtained by padding $\variable{randA}$ and $\variable{randT}$ with 0s to make them 16 bytes long. Essentially, the tag proves its knowledge of $\variable{authKey}$ to the owner in this step. 

\heading{Location reporting by the owner}
Once a connection has been established, the owner device periodically sends location reports to the server through an HTTPS POST request containing its location information---altitude, latitude, longitude, and timestamp---along with a list of $\variable{tileId}$s and corresponding authentication data---$\variable{randA}$, $\variable{randT}$, $\variable{sresT}$---for each connected tag.
We detail the request body in~\Cref{sec:http-request-bodies}.
Interestingly, the owner periodically uploads location reports for a Tile tag to the server even when it is connected to the tag.
This step is not necessary for the functionality of an OF protocol and poses privacy risks that we will discuss in~\secref{subsec:vulnsAgainstLocationPrivacy}.

\subsection{Tag-Finder-Server interactions}\label{subsec:tag-finder-server-interactions}

\begin{figure}[t] 
\centering
\small
\begin{tikzpicture}
\small{
\tikzstyle{arrow} = [->, thick]
\tikzstyle{block} = [rectangle, draw, fill=white, text centered, minimum height=2em, minimum width=2em]

\node (tile_owner) at (0,0) {
    \includegraphics[width=2.5cm]{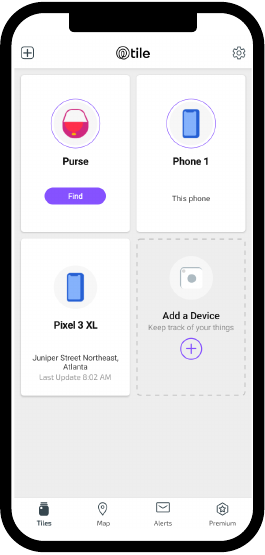}
};

\node (tile_tag) at (6,1) {
    \includegraphics[width=1.5cm]{tile.png}
};

\node (tile_server) at (6,-1.5) {
    \includegraphics[width=2cm]{service-provider.png}
};

\draw[arrow] (5.0, 1.5) -- (1.5, 1.5) node[midway, above, align=center] {1. BLE advertisements\\ containing $\variable{privateId}$s\\and static MAC addresses};

\draw[arrow] (1.5, -1.3) -- (5.0, -1.3) 
    node[midway, above, align=center] {2. Location data,\\ $\variable{privateId}$s, MAC addresses};
}
\end{tikzpicture}
\caption{Location reporting in the lost mode.}
\label{fig:location-reporting}
\end{figure}

We now outline the core of Tile's OF protocol.
Specifically, we describe the interactions between a finder, a tracker tag, and the server. 
We summarize this phase in~\Cref{fig:location-reporting}.

\heading{Tag-Finder interactions}
An activated tag rotates through $\variable{privateId}$s every 15 minutes regardless of its connection to the owner.
This means that Tile tags do not differentiate between lost and connected modes---they are \emph{always} findable.

The advertisements emitted by a tag contain the tag's rotating $\variable{privateId}$.
A finder/bystander device running the Tile app executes a service that periodically scans for BLE advertisements in the background.
The app filters BLE advertisements using the `FEED' service UUID to get advertisements from nearby Tile tags.
The finder device then parses the advertisement data and retrieves the tag's $\variable{privateId}$ and MAC address.
It checks if the $\variable{privateId}$ corresponds to any of its own tags.
If not, the finder adds the $\variable{privateId}$ and MAC address to a database of seen tags.

\heading{Finder-Server interactions}
Periodically, the finder device also runs a service that batches multiple scanned advertisements and computes its own location at the time.
It then uploads its location and the batched advertisements along with its own $\variable{clientID}$ to Tile's servers via an HTTPS POST request.
We describe the request body in~\Cref{sec:http-request-bodies}.
The uploaded advertisement data contains the \emph{static} MAC address of the advertising tag as well as the corresponding $\variable{privateId}$.
Finder devices also uniquely identify themselves as the user uploading reports to Tile's server.

\subsection{Owner-Server interactions}\label{subsec:owner-server-interactions}
When the owner wants to track their Tile, they log in and query the tiles/location/history/\{tileId\} endpoint to retrieve location history for their Tile with Id $\variable{tileId}$.
The owner includes its $\variable{user\_uuid}$ in the request header.
The server first verifies whether the requested $\variable{tileId}$ is registered to the requesting user's $\variable{user\_uuid}$. 
If the verification succeeds, the server returns the location data for the corresponding Tile as reported by finders; otherwise, it rejects the query.

\subsection{Scan and Secure}\label{subsec:scan-and-secure}
The Scan and Secure feature allows a user to trigger a scan for unknown Tiles or Tile-enabled devices that may be following the user.
Unlike similar features provided by other service providers like Apple and Google, this feature is not built into the operating system or available to non-Tile users. 
It is not even enabled in the app by default.
To run a scan, a user must manually trigger it by clicking the ``Scan and Secure'' button under the app's settings.
The scan takes approximately 10 minutes and requires the user to be physically moving while performing the scan and monitoring results on the screen. 

During the scan, the app performs six consecutive Bluetooth scans, filtering results based on the `FEED' service UUID.
It then parses the advertised data and retrieves the $\variable{privateId}$ from each advertisement.
For each $\variable{privateId}$, the app checks it against its local database of $\variable{privateId}$s to determine whether the $\variable{privateId}$ corresponds to a connected/paired tracker. 
If so, it categorizes the corresponding result as a ``Known Tile''.
All other advertisements are categorized as having come from an ``Unknown Tile''.
For unknown trackers, the app sends an HTTPS POST request to the Tile server containing the corresponding $\variable{privateId}$s for each scan.
We provide the format of the request body in~\Cref{sec:http-request-bodies}.

The server responds with a subset of the $\variable{privateId}$s that do not correspond to Tiles that are in the Anti-Theft mode.
We explain the Anti-Theft mode shortly.

Finally, the app then displays two lists: one for ``Known Tiles'', where it shows the user-assigned local names of connected trackers, and another for ``Unknown Tiles'', where it displays the $\variable{privateId}$s returned by the server along with the number of times (out of six) each $\variable{privateId}$ appeared.

\subsection{Anti-Theft mode}\label{subsec:antitheft-mode}
Anti-Theft mode allows users to make their Tile trackers invisible to the Scan and Secure feature. 
To enable this mode, users must verify their identity by providing a government-issued ID and live photos.
Tile relies on two third-party verification services, Berbix and Persona, to handle this process. 
Once verification is complete and the user agrees to Tile's terms, the server updates the user's Anti-Theft status to ``enabled''.

As a result, Tiles owned by that user are excluded from Scan and Secure results. 
In particular, when the server receives a list of $\variable{privateId}$s during a Scan and Secure scan, it filters out any $\variable{privateId}$s associated with Tiles in Anti-Theft mode, returning only those that do not have this setting enabled.

\subsection{Transfers and Unlimited Sharing}\label{subsec:tag-transfers-and-sharing} 
Tile provides a feature for users to transfer ownership of their trackers to another user and share their Tiles with other users. 

We discuss transfers first.
To initiate a transfer, the current owner selects the Tile they wish to transfer, navigates to ``More Options,'' chooses ``Transfer,'' and enters the recipient's email address. 
This action triggers an HTTPS POST request to the Tile production API server, which includes the recipient's email address. 
Upon successful transfer, the original owner receives an email notification, though the recipient is not explicitly notified. 
Instead, the transferred Tile automatically appears in the recipient's list of connected Tiles when they log in or reopen the Tile app. 
Notably, the server forwards the $\variable{authKey}$ from the original owner to the new recipient---there is no attempt to update the $\variable{authKey}$ after transfer.

Now we discuss Tile's ``Unlimited Sharing'' feature.
Tile allows users to share their Tiles with other users.
In order to share a Tile, a user selects the Tile they wish to share, chooses ``Unlimited Sharing'' and enters the email address of the new shared owner. 
Then the app sends an HTTPS request to the server containing the $\variable{tileId}$ of the Tile and the email address of the new owner.
The server associates the corresponding tag with the new owner and sends them the $\variable{authKey}$.
We describe the server's response body in~\Cref{sec:http-request-bodies}.

Unlike the transfer feature, neither the original owner nor the new shared owner is explicitly notified about the sharing of the Tile.
The tile appears on the new user's screen and is displayed as a shared tile. 
To stop sharing a Tile with a user, an owner selects the Tile on the app's home screen, chooses ``Unlimited Sharing'' option, and selects the corresponding shared user. 
This initiates an HTTPS DELETE request including the Tile's $\variable{tileId}$ and the shared owner's email. 
The tag is removed from the shared owner's home screen. 
Importantly, the $\variable{authKey}$ is not updated even after the original owner stops sharing their Tile with another user.

\subsection{Community information}\label{subsec:community-info}
\begin{figure}[t]
\centering
\small
\begin{tikzpicture}
\small{
\tikzstyle{arrow} = [->, thick]
\tikzstyle{block} = [rectangle, draw, fill=white, text centered, minimum height=2em, minimum width=2em]

\node (tile_owner) at (0,0) {
    \includegraphics[width=4cm]{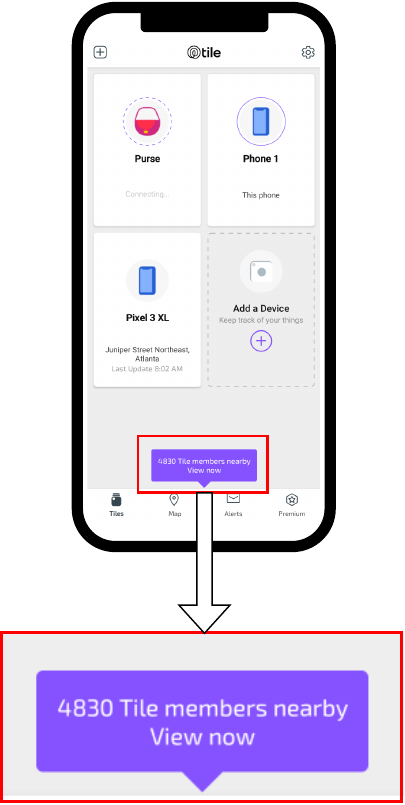}
};

\node (tile_owner_2) [right=of tile_owner, xshift=-0.9cm] {
    \includegraphics[width=4cm]{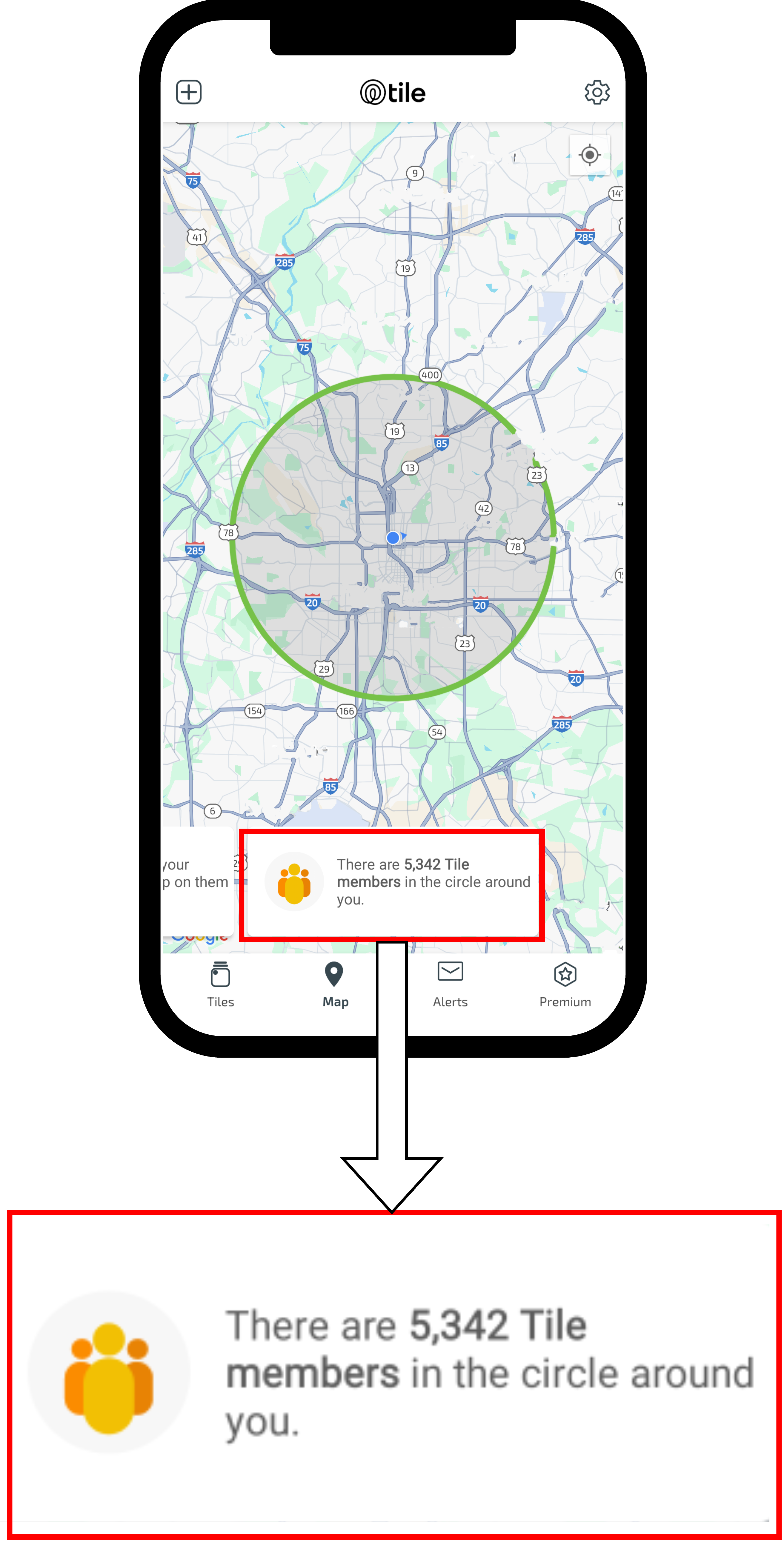}
};

}
\end{tikzpicture}
\caption{Tile's community information feature.}
\label{fig:community-info}
\end{figure}

Tile has a feature called Community Information that is not directly part of its OF protocol.
This feature displays the number of Tile users in a 5-mile radius around the owner.
We show a screenshot of the feature in \Cref{fig:community-info}.
When a user opens the Tile app, their client sends an HTTPS GET request to the Tile API server at /api/v1/community/stats, including the latitude and longitude of the client's current location. 
The server responds with the number of Tile users in the 5-mile radius around the client. 
We describe the server's response format in~\Cref{sec:http-request-bodies}.
We note that this feature is not useful for the functionality of the OF network.

\subsection{Account deletion}\label{subsec:account-deletion}
To delete a Tile account, the user must navigate to the Settings menu and select `Delete Account' under `Manage Account'.
The process requires the user to enter their password and acknowledge that their Tiles will no longer be functional and cannot be reactivated, and that account deletion is a permanent action. 
Additionally, users must confirm that, although their account details and personal information will be removed from Tile's servers, ``Tile may need to retain certain data'' per Tile’s privacy policy.
Finally, the user is required to type in `DELETE' to confirm account deletion. 
This triggers an HTTPS DELETE request to the Tile API server at /api/v1/users/$\variable{user\_uuid}$.
Upon deletion, the server sends an HTTP 202 Accepted status message.
\section{Privacy and security analysis}\label{sec:security-analysis}
We now detail vulnerabilities that we identified against Tile's OF protocol, classifying them into those that (1) allow the platform server to persistently learn the location information of owners and finders by breaking the location indistinguishability property (\secref{subsec:vulnsAgainstLocationPrivacy}), (2) compromise owner privacy by breaking the tag indistinguishability property (\secref{subsec:vulnsAgainstOwnerPrivacy}), (3) enable stalking by breaking the tag detetctability property (\secref{subsec:vulnsAgainstTagDetectability}), (4) allow the framing of honest users by breaking the accountability mechanism (\secref{subsec:vulnsEnablingstalking}), and (5) allow the circumvention of the Anti-theft mode (\secref{subsec:vulnsAgainstAntitheft}).

To enable independent verification, we describe the experimental methodology we used to confirm the practical exploitability of the vulnerabilities we report. 

\subsection{Violating location indistinguishability}\label{subsec:vulnsAgainstLocationPrivacy}

We outlined in~\secref{subsec:sec-props-and-threat-models} that an OF protocol should ensure location indistinguishability against the platform server. 
However, vulnerabilities in Tile's implementation undermine this property, allowing (1) Tile's server to run a mass surveillance network and (2) any third-party attacker with the Tile application to systematically deanonymize and track Tile users. 
These issues transform Tile's infrastructure into a global tracking network.

\heading{1. Tile's servers continuously collect unencrypted location information of its users}
Tile's protocol allows the platform server to persistently collect unencrypted location information about all Tile users. 
For reports submitted by finder devices, this also includes the static MAC address of the corresponding tag.
Other service providers (like Apple and Google) encrypt the location using the destination’s public key and only allow their tracker tags to be discoverable by non-owners when in the “lost” mode i.e., the trackers are only detectable by finder devices when disconnected from the owner’s device. 

This compromises location indistinguishability and directly contradicts the claim made in Tile’s Security and Privacy Policy~\cite{TilePrivacyPolicy}, which tells users that they are the \emph{only} ones who can access the location of their device(s) and Tile(s). 

We verified our claim as part of the network traffic analysis we performed to reverse engineer the Android application. 
While intercepting the HTTPS traffic between the Tile Android client and Tile's servers, we confirmed that location reports were transmitted over TLS in plaintext (see~\Cref{sec:http-request-bodies}).

\heading{2. The Community information feature compromises users' location privacy}
The Community Info feature may allow an attacker to deanonymize the identities of Tile users.
Community stats display the number of Tile users in a 5-mile radius. 
An attacker could query the API endpoint on various latitude and longitude values and use these aggregate values to deanonymize users. 
We note that attacks in this setting have been well documented in the literature~\cite{10.1145/3038912.3052620,Pyrgelis2017WhatDT}.

The service provider could mitigate this attack by rate limiting these queries or obfuscating the results. 
However, implementing rate limiting is non-trivial since an adversary could just create multiple accounts and make one query using each account.
Due to ethical and legal concerns, we chose not to send any irregular traffic to Tile’s servers.
Hence, we were unable to confirm if this vulnerability is exploitable.

\subsection{Violating tag indistinguishability}\label{subsec:vulnsAgainstOwnerPrivacy}

If an OF protocol does not guarantee tag indistinguishability against passive RF adversaries, then it can be transformed into a pervasive surveillance infrastructure. 
To motivate this, consider a situation in which an attacker (e.g., a government agency, cell service provider, or app developer) with an RF receiver is able to uniquely identify a BLE advertisement as having been broadcast by their target's tag.
The attacker could deploy RF receiver networks---for example, by using an app or SDK with fine-grained Bluetooth access---to log target devices' advertisements. 
By correlating these identifiers across time and geographic location, particularly in high-traffic areas like transit hubs or city centers, the attacker could construct detailed movement profiles of individuals without their knowledge or consent.

As outlined in~\secref{sec:tiles_claims_of_security}, Tile's privacy policy~\cite{TilePrivacyPolicy} explicitly claims that all communications over their network are anonymous and that no one can learn where others' Tiles are.
The following four vulnerabilities not only break the tag indistinguishability property as defined in the literature but also directly contradict Tile's claimed privacy guarantees. 

\heading{1. The Tile Mate 2022 broadcasts static MAC addresses}
MAC addresses broadcast by the Tile device are static. 
This allows a passive RF adversary to record MAC address broadcast by a Tile Mate and use them to identify and monitor the user's location. 
As a result, the work put into rotating $\variable{privateId}$s in Bluetooth advertisements does not guarantee any privacy/anonymity to the owner of a Tile Mate.
 
To confirm that the Tile Mate 2022 broadcasts static MAC addresses, we recorded all BLE communication between our test phone and nearby peripherals.
Over multiple iterations spanning several days, we captured BLE advertisement packets from the Tile Mate 2022 and analyzed the corresponding log files.
We consistently observed the same static MAC address in all advertisement packets emitted from the device.

\heading{2. $\variable{privateId}$s do not guarantee tag indistinguishability even if MAC addresses are randomized}
There are 8640 unique $\variable{privateId}$ values corresponding to a particular Tile/Tile-enabled device. 
A device rotates its $\variable{privateId}$ every 15 minutes. 
Instead of regenerating a new set of $\variable{privateId}$ values once the protocol uses all 8640 values, Tile begins re-using the same $\variable{privateId}$s. 
As a result, TileIDs are repeated every 90 days. 
A passive RF adversary could record all the $\variable{privateId}$s emitted by a tag over 90 days and use these values to uniquely identify the tag even if it has randomized MAC addresses. 

To verify our claim, we conducted the following experiment.
We programmatically logged in to our test user account on the Tile Android app and retrieved the $\variable{authKey}$ associated with a specific Tile Mate 2022 registered to the account. 
The algorithm used for generating all (8640) unique $\variable{privateId}$ values is detailed in~\secref{subsec:tag-activation}. 
We used the algorithm to generate all 8640 unique values for our test device, 
and recorded the BLE advertisements emitted by the Tile Mate over 90 days by scanning for BLE advertisements. We compared the values captured with the set of previously generated 8640 $\variable{privateId}$s, and 
observed that the $\variable{privateId}$s repeated after 90 days, confirming that Tile reuses the same set of values. 

This reuse pattern---together with the collision resistance property of the $\variable{HMAC\mbox{-}SHA256}$ hash function used to generate the $\variable{privateId}$s--- allows an adversary to link $\variable{privateId}$s over time and track the device.
This is a realistic adversary: Individuals are likely to visit the same location at the same time of day (e.g., one's home or office) over long periods of time, and Tile devices are intended to be used for over a year. 

\heading{3. The Scan and Secure feature does not guarantee the privacy of the owners of detected Tiles}
Tile claims that the results of Scan and Secure present a ``unique and encrypted identifier'' that ensures the privacy of the corresponding owner~\cite{TileScanAndSecureFAQ}.
We find that this claim is both technically incorrect and substantively misleading. 

As described in~\secref{subsec:scan-and-secure}, Tile's Scan and Secure feature's results contain the $\variable{privateId}$s that appear during the six Bluetooth scans (as long as they do not correspond to a Tile in the Anti-theft mode), even if the $\variable{privateId}$ was recorded only once. 
Contrary to Tile's claim, the identifiers in the Scan and Secure result ($\variable{privateId}$s) are pseudorandom, not encrypted. 

Furthermore, the identifier does not provide privacy to the Tile owner, especially in the case of Tile Mate 2022, as the MAC address is static.
The static MAC address is embedded in the advertisements recorded during the scan and allows the scanner to permanently fingerprint a device.
This architectural flaw not only invalidates Tile's privacy guarantees but creates systemic surveillance risks---allowing any scanner to deanonymize innocent bystanders whose Tiles were incidentally detected during scans. 

\heading{4. The Transfer/Unlimited Sharing features compromise the $\variable{authKey}$}
As discussed in~\secref{subsec:tag-transfers-and-sharing}, the transfer and/or sharing process does not regenerate or change the $\variable{authKey}$---meaning that both users gain access to the same $\variable{authKey}$ used by the tag. 
Since $\variable{privateId}$s are deterministically derived from the $\variable{authKey}$, either recipient can generate all $\variable{privateId}$s associated with the shared or transferred tag.

While transferring ownership is a permanent action, sharing a Tile can be revoked.
However, even if a shared user’s access is later revoked, they can still leverage the $\variable{authKey}$ or derived $\variable{privateId}$s to track the original owner of the Tile.

This attack is particularly relevant in cases of Intimate Partner Violence (IPV), where Bluetooth trackers have been reported to aid the perpetrator. 
In this setting, an abuser might gain temporary access to their victim's phone, transfer or share the victim's Tile to themselves, obtain the $\variable{authKey}$, and then transfer the Tile back or revoke their own shared access. 
With the $\variable{authKey}$ in hand, they could track/stalk their victim by generating all $\variable{privateId}$ values, then using RF receivers to track the victim using these values.

\subsection{Violating tag detectability}\label{subsec:vulnsAgainstTagDetectability}

The tag detectability property requires that unwanted tracking attempts be reliably detected by potential victims.
With the growing use of Bluetooth-based trackers for stalking, this property is imperative in OF systems.

We identify fundamental flaws in Tile's design that violate this property, allowing malicious actors to evade detection and stalk users. 
The tag detectability property is trivially broken for the majority of users without Tile's app: To detect a malicious actor, a victim must install Tile's application, creating additional barriers to discovery. 
And, even if the app is installed, Scan and Secure has a number of usability issues: Scan and Secure is not a background process, and must be triggered manually in a sub-menu on the app. 
Worse, the scan duration is ephemeral, lasts about 10 minutes, and will not continue once the user navigates away.
As we will discuss in ~\secref{sec:discussion}, this reactive, opt-in approach contrasts sharply with the continuous and automatic background scanning implemented in competing systems. 

Furthermore, Tile's Anti-Theft mode may be used to trivially violate tag detectability---Tiles in the Anti-Theft mode will not be detected by its Scan and Secure feature (unless, as we will discuss in~\secref{subsec:vulnsAgainstAntitheft}, the user uses a modified app).  
Tile intentionally made this trade-off to support their anti-theft use case which is in direct tension with the tag detectability property. Though such tensions may be theoretically resolvable through careful design, Tile’s implementation worsens the system’s already weak detectability, and remains an open area of research.

\subsection{Violating framing resistance}\label{subsec:vulnsEnablingstalking}

As discussed in~\secref{subsec:sec-props-and-threat-models}, one of the main technical challenges of implementing an accountability mechanism for OF networks is achieving framing resistance. 
Below we describe replay attacks that exploit Tile's protocol to frame honest users.

\heading{1. Scan and Secure results are spoofable}
As we outlined in \secref{sec:protocol-description}, the Scan and Secure feature initiates 6 consecutive Bluetooth scans to detect advertisements from Tile or Tile-enabled devices, requiring the user to be in motion.
Tile then extracts the $\variable{privateId}$s embedded in the scanned advertisements and forwards the discovered $\variable{privateId}$s to the server.
The server returns only those $\variable{privateId}$s that are not associated with Tiles in Anti-Theft mode.

An attacker with a compromised $\variable{authKey}$ (e.g., as described in \secref{subsec:vulnsAgainstOwnerPrivacy}) could execute a derive-then-replay attack to frame the compromised user.
It would first derive $\variable{privateId}$s from the compromised $\variable{authKey}$, then broadcast these values at chosen locations. 
If a user runs the Scan and Secure feature in that location, the app will display a false positive, suggesting the presence of a Tile where none may exist. 
Indeed, there is no way to prove that the detected $\variable{privateId}$ was emitted by a legitimate Tile device.

\heading{2. Users can be framed for misconduct}
We introduced framing resistance in \secref{subsec:sec-props-and-threat-models}.
We claim that both active RF adversaries and the platform server can frame honest Tile users for misconduct. 

An active RF adversary would first passively listen for and collect all the $\variable{privateId}$s broadcast by an honest Tile.
It would then simply replay the $\variable{privateId}$s of the honest tag at an arbitrary location.
Tile's Scan and Secure would record these replayed $\variable{privateId}$s and display false positives to a user.
Finally, Tile's server has the $\variable{authKey}$ for all Tile devices, and can therefore frame any user. 

\subsection{Circumventing the Anti-Theft mode}\label{subsec:vulnsAgainstAntitheft}

Tile uniquely markets its Anti-theft mode as a solution for theft protection, claiming tags in the Anti-theft mode cannot be discovered via Scan and Secure. 
However, our analysis reveals that this protection is superficial---while tags in Anti-Theft mode are excluded from the user interface, the system \textit{still} records and transmits their $\variable{privateId}$s during scanning operations. 
This design choice allows for trivial circumvention that fundamentally undermines Tile's advertised guarantees.

\heading{Anti-Theft mode can be circumvented by Tile users}
Our analysis in \secref{subsec:scan-and-secure} reveals the Scan and Secure feature records \textit{all} observed $\variable{privateId}$s, including those from tags in the Anti-Theft mode. 
The server merely filters these tags from the displayed results.
A user with a modified app can easily circumvent the Anti-Theft mode by displaying all $\variable{privateId}$s recorded during the scan. 

To verify that Tile's Anti-Theft Mode can be circumvented, we conducted an experiment using two test accounts (A and B) on our test Android device.
We enabled the Anti-Theft Mode in account A, and while logged in to account B on a different device, we ran Scan and Secure while in the vicinity of tags registered under account A.
While analyzing network traffic from B, we confirmed that the HTTPS POST request sent to Tile's server contained $\variable{privateID}$s emitted by A's anti-theft-mode-enabled tag.

\section{Discussion}\label{sec:discussion}

\begin{table*}[t]
    \centering
    \caption{Privacy properties of different OF networks.}
    \begin{tabular}{lccccc}
        \toprule
        \textbf{Privacy Property} & \textbf{Tile} & \textbf{Apple} & \textbf{Google} & \textbf{Samsung} & \textbf{RFC (DULT)} \\
        \midrule
        Tag indistinguishability & $\times$ & \checkmark & \checkmark & $\times$ & \checkmark \\
        Location indistinguishability & $\times$ & \checkmark & \checkmark & \checkmark\textsuperscript{*} & \checkmark \\
        Tag broadcasts randomized MAC addresses & $\times$\textsuperscript{\ddag} & \checkmark & \checkmark & \checkmark & \checkmark \\
        \bottomrule
        \label{table:3}
    \end{tabular}

    \vspace{1pt}
    
    {\footnotesize \textsuperscript{*}Samsung was reported insecure by Yu et al.~\cite{yu2022privacy}, but has been fixed since to provide location indistinguishability in modern phones, and optionally also for its custom tags.
    \textsuperscript{\ddag} Result for Tile Mate 2022.}
\end{table*}

\begin{table*}[t]
    \centering
    \caption{Antistalking and accountability properties of different OF networks.}
    \begin{tabular}{lccccc}
        \toprule
        \textbf{Anti-Stalking Property} & \textbf{Tile} & \textbf{Apple} & \textbf{Google} & \textbf{Samsung} & \textbf{RFC (DULT)} \\
        \midrule
        Tag detectability (stalking alerts) & $\times$ & \checkmark & \checkmark & \checkmark & \checkmark \\
        Tag detection triggered automatically by iOS/Android & $\times$ & \checkmark & \checkmark & \checkmark\textsuperscript{*} & \checkmark \\
        Attempted accountability & \checkmark & $\times$ & $\times$ & $\times$ & $\times$ \\
        Framing resistant & $\times$ & \checkmark & \checkmark & \checkmark & \checkmark \\
        \bottomrule
        \label{table:4}
    \end{tabular}
    
    \vspace{1pt}
    {\footnotesize * This feature is only available in Samsung phones, through Samsung's proprietary SmartThings app. In Samsung phones, it is enabled by default.}
\end{table*}

\heading{A comparison of the security of major OF networks}
To contextualize the security implications of Tile's system, we compare its protocol with three competing players in the OF landscape---Apple, Google, and Samsung---as well as the IETF's DULT draft standard.
We evaluate these protocols with respect to the security properties defined in~\secref{subsec:sec-props-and-threat-models}, and divide them across their privacy properties (\Cref{table:3}) and their anti-stalking and accountability features (\Cref{table:4}).

We start by discussing privacy properties.
Each of the three major vendors and the DULT specification implements end-to-end encryption for location data under keys only accessible to the reporting finder and owner of the lost tag. 
Crucially, platform servers learn nothing about user locations---a fundamental privacy guarantee that Tile fails to provide. 
Previously, Samsung was shown to be inadequate on both accounts~\cite{yu2022privacy}---tag indistinguishability and location indistinguishability---but has since updated its protocol to improve tag indistinguishability and optionally provide location indistinguishability if a user explicitly enables it in their system settings.
Tile is alone in its continued use of static MAC addresses, a choice which renders much of Tile's already limited defenses against passive RF adversaries moot.  

Next, we consider antistalking and accountability properties.
For the former, we evaluate whether providers' antistalking algorithms successfully detect rogue tags as well as whether they are always enabled by default and alert users automatically.
All service providers except Tile have implemented antistalking algorithms that guarantee tag detectability at the operating system level, ensuring that these scans always run in the background and alert the user automatically. 
Samsung limits this protection to its own devices, while Apple, Google, and DULT extend coverage to all iOS and Android smartphones. 
Tile's reliance on manual, user-initiated scans creates dangerous detection gaps. 
Naturally, since Tile operates only at the application level, it lacks OS-level privileges and cannot support background scanning unless integrated into Apple’s or Google’s protocols.

Tile is the only deployed OF network that attempts to provide accountability. 
However, as demonstrated in~\secref{sec:security-analysis}, its implementation is naive and introduces framing vulnerabilities absent in other systems.
Apple and the DULT specification allow a non-owner with physical access to a tag to query partial account information of the owner. 
This feature is optional; there is nothing cryptographically tying this information to advertisements, and it does not appear to have much use as an accountability mechanism.

 \heading{The need for increased transparency}
All major OF systems, including those by Apple, Samsung, and Tile, required reverse engineering for initial security analyses. 
These analyses exposed various vulnerabilities and limitations across the systems.
Increased transparency will enable security researchers to efficiently review and identify potential weaknesses \textit{before} the system is deployed.
Transparency can also foster trust between users and service providers and better allow users to understand the inherent risks of the systems they depend on.

\heading{Security guarantees provided by service providers should be well-defined}
Our work demonstrates that many of Tile's security claims were (a) incorrect (user anonymity, location privacy), (b) insinuated, but substantively wrong (detection of rogue tags using the Scan and Secure feature is valid/unspoofable), (c) correct, but vulnerable to an active attacker (accountability for the abuse of the Anti-Theft mode). 
These findings highlight a broader imperative: service providers should offer clear and accessible definitions for the security properties they attempt to guarantee, alongside a well-defined threat model outlining the conditions under which these properties hold. 
Without this context, users have little chance of making informed choices about their own security and privacy needs.

\heading{Cryptographic accountability in OF networks}
The design of Tile's accountability mechanism is flawed, subvertible, and allows for framing attacks, but these design issues are informative. 
Tile's protocol lacks clarity on what kind of evidence can be produced to implicate a user, who can access this evidence, and under what circumstances this evidence holds validity. 
The challenge of incorporating accountability mechanisms into OF networks remains a worthwhile research direction.
Future work that establishes strong and practical definitions of accountability can enable victims to work with service providers or trusted third parties to identify misuse, prove harm, and, ultimately, disincentivize stalking and abuse.

\heading{Fundamental tensions in OF security goals}
The security notions for offline finding networks appear to balance three seemingly conflicting goals: (1) tag indistinguishability, which prevents an attacker from linking Bluetooth advertisements to a particular tag; (2) tag detectability, which requires that a victim that has seen a sufficient number of advertisements from the same tag should be able to link them to form an alert; and (3) accountability, which requires that if a user detects a malicious tag, then they should be able to collaborate with appropriate authorities to punish the (correct) tag's owner. 
Current implementations prioritize different subsets of these properties at the expense of others. 
Further systematization, formalization of tradeoffs, and development of cryptographic systems that attempt to satisfy all three requirements remain open research problems worthy of future work.

\section{Conclusion}\label{sec:conclusion}
We reverse engineer Tile's OF protocol and provide an in-depth security and privacy analysis.
Tile's unique Anti-Theft mode intentionally weakens anti-stalking safeguards, and includes an accountability mechanism for penalizing abusers of the feature.
We identify several attacks that compromise the privacy of Tile's users and find that the Scan and Secure anti-stalking feature is inadequate for protecting individuals from being tracked by Tile devices.
Additionally, we find that the Anti-Theft mode is easily circumventable, and that a number of adversaries can subvert the accountability mechanism to frame honest users.
Ultimately, argue that introducing to accountability OF networks is an open problem, advocate for future research in implementing useful solutions that balance privacy and accountability.

\clearpage
\section{Ethical Considerations}

\heading{Process}
To ensure that our actions did not impact Tile's servers, we sent no unnatural traffic and only performed attacks against our own devices and infrastructure. Any communication with Tile's servers were performed following their protocol for normal use.

\heading{Disclosure}
We disclosed our findings to Tile on November 13, 2024, by contacting CEO Chris Hulls (chris@life360.com) and the support team (support@life360.com), as a direct vulnerability disclosure channel was not available. 
We committed to adhere to the industry standard of 90 days (through February 11, 2025) before publicly disclosing our findings and offered remediation assistance. 
Tile acknowledged the vulnerabilities and engaged in dialogue until February 4, 2025, after which communications ceased. 
Following an additional notice on February 5 declaring our intent to publish, we extended the embargo period by an additional 60 days (until April 7, 2025) to give them sufficient time to address issues.

\heading{Mitigations}
As part of our disclosure, we offered to provide Tile with recommendations for mitigating the vulnerabilities we identified. 
Some mitigations are feasible via firmware updates: for location indistinguishability, we suggest adopting end-to-end encryption of location data, as implemented by Samsung in response to a related issue; for tag indistinguishability, we recommend replacing static with randomized MAC addresses, deriving $\variable{privateId}$s in real time using an incrementing counter to avoid long-term reuse, and updating the $\variable{authKey}$ upon tag transfer or sharing to reduce linkability.
Other mitigations require more substantial architectural changes: since Tile’s application lacks the OS-level privileges needed for continuous scanning (unlike Apple's and Google’s implementations), rebuilding the application around OS-native frameworks is required to improve tag detectability. 
Finally, with the above mitigations in place, Tile could, in principle, strengthen framing resistance by verifying whether a $\variable{privateId}$ could have been generated by any valid $\variable{authKey}$ for the corresponding timestamp, although this approach would be computationally inefficient in practice.

\bibliographystyle{plain}
\bibliography{references}

\clearpage

\appendix
\section{HTTP request/response bodies}\label{sec:http-request-bodies}

\heading{Body of the HTTPS request sent by the owner device to the server during the secret key establishment process of the tag activation phase~\secref{subsec:tag-activation}}

\begin{center}
\small
\begin{verbatim}
    {     
        "tile_uuid": tileId,
        "name": "Mate",
        "rand_a": rand_a,
        "rand_t": rand_t,
        "sres_t": sres_t,
        "hw_version": 24.00,
        "model": TILE 24.00,
        "firmware_version": 48.04.16.0
    }
\end{verbatim}
\end{center}

Here, the corresponding values for \texttt{tile\_uuid}, \texttt{name}, \texttt{hw\_version}, \texttt{model}, and \texttt{firmware\_version} are obtained during TDI and \texttt{rand\_a}, \texttt{rand\_t}, and \texttt{sres\_t}, are obtained during tag authentication. 
\vspace{10pt}

\heading{Location reports submitted by an owner for connected tags}  

\begin{center}
\small
\begin{verbatim}
    {"updates": [{
        "record_id": x,
        "location": {
            "altitude": x,
            "latitude": x,
            "longitude": x,
            "timestamp": x,
            ...},
        "tiles": [{
            "connected_auth_data": {
                "rand_a": rand_a,
                "rand_t": rand_t,
                "sres_t": sres_t,
                "tile_uuid": <8-byte UUID>},
            "discovery_timestamp": x,
            "record_id": x
            },
            {
            "client_data": {"tile_uuid": x},
            "discovery_timestamp": x,
            "record_id": x
            }]
        },
        ...
    ]}
\end{verbatim}
\end{center}
Here, the corresponding values for \texttt{rand\_a}, \texttt{rand\_t}, and \texttt{sres\_t} are obtained during tag authentication, and the location values include the owner's current location.
\vspace{10pt}

\heading{Location reports submitted by a finder for lost tags}
\begin{center}
\small
\begin{verbatim}
    {"updates": [{
        "record_id": x,
        "location": {
            "altitude": x,
            "latitude": x,
            "longitude": x,
            "timestamp": x,
            ...},
        "tiles": [{
            "advertised_service_data": {
            "mac_address": MAC address,
            "payload_service_data": privateId,
            ...},
            "discovery_timestamp": x,
            "record_id": x
            },
            {
            "client_data": {"tile_uuid": x},
            "discovery_timestamp": x,
            "record_id": x
            }]
        },
        ...
    ]}
\end{verbatim}
\end{center}

Here \texttt{payload\_service\_data} includes the $\variable{privateId}$ broadcast by the lost tag, and \texttt{mac\_address} is the tag's static MAC address. 
\vspace{10pt}

\heading{Request sent by a Scan and Secure client to the server after completing the scan}

\begin{verbatim}
[
    {"privateIds": [x1, x2, ...]},
    {"privateIds": [x1, x2, ...]},
    {"privateIds": [x1, x2, ...]},
    {"privateIds": [x1, x2, ...]},
    {"privateIds": [x1, x2, ...]},
    {"privateIds": [x1, x2, ...]}
]
\end{verbatim}
Here, each \texttt{xi} is a $\variable{privateId}$ value seen during the corresponding scan.
\vspace{10pt}

\heading{The server's response for the Unlimited Sharing feature} 
\begin{verbatim}
    "result":{
        "tileType":"TILE",
        "tile_uuid": x,
        "user_uuid": x,
        "other_user_uuid":x,
        "other_user_email":x,
        ...
        }
\end{verbatim}
Here \texttt{other\_user\_uuid} and \texttt{other\_user\_email} correspond to the shared owner's details. 
\vspace{10pt}

\heading{The server's response for the Community Information feature}
\begin{verbatim}
 {...
  "timestamp_ms": x,
  "result_code": 0,
  "result": {
    "timestamp": x,
    "center_latitude": lattitude,
    "center_longitude": longitude,
    "center_radius": 5.0,
    "tilers_around": numberOfTileUsers,
    "display_tilers_around": true,
    "display_tiles_found": false, 
    ...
  }
}   
\end{verbatim}

Here the latitude and the longitude are the client's current location coordinates, the center radius is always set to 5 miles, and the \texttt{tilers\_around} value represents the number of Tile users in the 5-mile radius around the client. 
\section{Details of Tile's protocol}

\subsection{Tile Device Information}\label{sec:char-uuids}
The owner requests the $\variable{tileId}$, $\variable{model}$, $\variable{firmware}$, and $\variable{hardware\_version}$ from the tag over the custom Tile Device Information (TDI) service (UUID 180A). 
The corresponding characteristic UUIDs for these variables are defined in~\Cref{table:1}.

\begin{table}[h]
\centering
\begin{tabular}{c c} 
 \hline
 Characteristic UUID & Value \\ [0.5ex] 
 \hline
 9d410007-35d6-f4dd-ba60-e7bd8dc491c0 & Tile Id  \\
 00002a24-0000-1000-8000-00805f9b34fb & Model number \\
 00002a26-0000-1000-8000-00805f9b34fb & Firmware \\
 00002a27-0000-1000-8000-00805f9b34fb & Hardware version  \\ [1ex] 

\end{tabular}
\caption{Characteristic UUIDs and corresponding values shared between the tag and the user under the $\mathsf{devInfoService}$ (UUID 180A) during activation.}
\label{table:1}
\end{table}

The $\variable{tileId}$ is an 8-byte static identifier of the tag that is derived from its static MAC address, the $\variable{model}$ is a 10-character alphanumeric identifier for the tracker of the form ``xxxx yy.yy'' where the first 4 characters constitute the vendor identifier assigned to a vendor by Tile and the last 5 characters stand for the model number, the value $\variable{firmware}$ is a 10-character numeric value of the form ``xx.xx.xx.x'' and represents the firmware version, and $\variable{hardware\_version}$ is a 5-character numeric value of the form ``xx.xx'' and represents the hardware version.

\subsection{Details of tag-owner interactions}\label{sec:near-mode-tag-owner-interactions}
\heading{The Tile Over-the-Air protocol}
Once a Tile tracker has been activated, communications with it use the \underline{M}ulti-\underline{E}nd\underline{p}oint \underline{T}ile \underline{O}ver-the-\underline{A}ir (MEP TOA) protocol.
MEP TOA is executed using two characteristics, MEP\_TOA\_CMD and MEP\_TOA\_RSP, defined under the `FEED' service. 
The corresponding character UUIDs are given in~\Cref{table:2}.
The former is used by the owner device to send messages to the tracker and
the latter is used by the tracker to send messages to the owner device.

\begin{table}[t]
\centering
\begin{tabular}{c c} 
\hline
 Characteristic Name & Characteristic UUID \\ [0.5ex] 
 \hline
 MEP\_TOA\_CMD & 9d410018-35d6-f4dd-\\
 & ba60-e7bd8dc491c0 \\
 MEP\_TOA\_RSP & 9d410019-35d6-f4dd- \\ 
 & ba60-e7bd8dc491c0

\end{tabular}
\caption{UUIDs for MEP\_TOA\_CMD and MEP\_TOA\_RSP characteristics used MEP TOA communications.}
\label{table:2}
\end{table}

Communications using MEP TOA can be connectionless or connected.
Connectionless communications are identified by a random 4-byte sequence called the  $\variable{toaToken}$.
Near-mode interactions between the owner and the tracker are over a connected channel.
We now describe the steps involved in establishing a connected channel.

\begin{figure}[t]
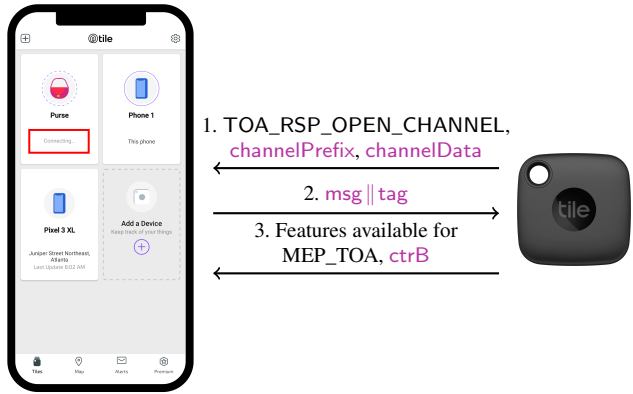

\centering
\footnotesize{
\begin{tikzpicture}
\tikzstyle{arrow} = [->, thick]
\tikzstyle{block} = [rectangle, draw, fill=white, text centered, minimum height=2em, minimum width=2em]

\node (tile_tag) at (6.3,-0.15) {
    \includegraphics[width=1.5cm]{tile.png}
};

\node (tile_owner) at (0,0) {
    \includegraphics[width=2.5cm]{tile_connect.pdf}
};

\draw[arrow] (5.3, 0.4) -- (1.5, 0.4) node[align=center, midway, above] {1. $\mathsf{TOA\_RSP\_OPEN\_CHANNEL}$,\\ $\variable{channelPrefix}$, $\variable{channelData}$};
\draw[arrow] (1.5, -0.2) -- (5.3, -0.2) node[midway, above] {2. $\variable{msg \concat tag}$};
\draw[arrow] (5.3, -1) -- (1.5, -1) node[align=center, midway, above] {3. Features available for\\ MEP\_TOA, $\variable{ctrB}$};
\end{tikzpicture}
}
\caption{The authentication protocol used by the owner device to authenticate itself to the Tile tracker. The variables $\variable{channelPrefix}, \variable{channelData}, \variable{msg}, \variable{tag}$, and $\variable{ctrB}$ are defined under owner authentication.}
\label{fig:owner-auth}
\end{figure}

\heading{Establishing a connected channel: Owner authentication}
The second part is owner authentication.
We summarize the steps involved in owner authentication in~\Cref{fig:owner-auth}.
This step begins once the Tile server sends a TOA\_OPEN\_CHANNEL message to the tracker containing the $\variable{channelPrefix}$ and $\variable{channelData}$ to be used for communications over the connected channel.
The owner first computes a key that we call the $\variable{tagKey}$ using the $\variable{authKey}$ as follows.
Here $\variable{randA}$ is the value used in the tag authentication step and $\variable{toaToken}$ is the identifier for the current connectionless channel.

\begin{center}
   {
   $\variable{seed} \gets \variable{randA} \concat \variable{channelData} \concat \variable{channelPrefix} \concat \variable{toaToken}$\\
   $\variable{tagKey} \gets \variable{HMAC\mbox{-}SHA256}(\variable{authKey}, \variable{seed})[0:128]$}
\end{center}

Then the owner authenticates itself to the tracker tag by sending a fixed message $\variable{msg = [0x12, 0x13]}$ along with a MAC tag over the message generated using $\variable{tagKey}$ as follows. 

\begin{center}
   {
   $\variable{seed} \gets \variable{ctrA} \concat 1 \concat \variable{msgLen} \concat \variable{msg}$\\
   $\variable{tag} \gets \variable{HMAC\mbox{-}SHA256}(\variable{tagKey}, \variable{seed})[0:32]$}
\end{center}
Here $\variable {msgLen}$ is the length of the message and $\variable{ctrA}$ is a counter maintained by the owner which is initialized to 0 and incremented each time a message is sent to the tracker.  
The tracker verifies the tag against the message by locally deriving $\variable{tagKey}$ from $\variable{authKey}$.

If the verification is successful, the tag responds with the features available for MEP\_TOA and the counter $\variable{ctrB}$.
The Tile Mate 2022 offers features such as ringing the Tile, (reverse) ringing the owner's phone.
After this step, the owner and the tracker have established a connected channel.
Each message the owner sends to the tag is authenticated with a MAC tag over the message generated using $\variable{tagKey}$ and $\variable{ctrA}$.

\heading{Ringing the tracker}
An owner can ring their Tile device by tapping on the corresponding tag on the app and clicking the ``Find'' button on the tag's screen.
This will use the song characteristic (9d410002-35d6-f4dd-ba60-e7bd8dc491c0) to play a song on the selected Tile tag.

\heading{Reverse ringing the phone}
The Find My Phone feature allows a Tile to ring the owner's phone using the reverse ring characteristic (9d410000-35d6-f4dd-ba60-e7bd8dc491c0) by double-pressing the button on the tag.
The owner can tap the ``Found it'' button on the notification to stop the ring.

\end{document}